\newcommand{\Msun}{{\rm M}_\odot}
\newcommand{\Mearth}{{\rm M}_\oplus}

\documentclass[12pt,preprint]{aastex}

\begin{document}
\title{Characterization of the Crab Pulsar's Timing Noise}
\author{D. M. Scott \altaffilmark{1}, M. H. Finger \altaffilmark{1} and 
C. A. Wilson} 
\affil{National Space Science and Technology Center, Huntsville,AL 35805}
\altaffiltext{1}{Universities Space Research Association}
\email{Matthew.Scott@nsstc.nasa.gov}

\begin{abstract}
We present a power spectral analysis of the Crab pulsar's timing noise, mainly 
using radio measurements from Jodrell Bank taken over the period 1982-1989,
an interval bounded by sparse data sampling and a large glitch. 
The power spectral analysis is complicated
by nonuniform data sampling and the presence of a steep red power 
spectrum that can distort power spectra measurement by causing severe 
power ``leakage''. We develop a simple windowing method for computing red 
noise power spectra of uniformly sampled data sets and test it on 
Monte Carlo generated sample realizations of red power-law noise. We 
generalize time-domain methods  of generating power-law red noise with even 
integer spectral indices to the    case of noninteger spectral indices. The Jodrell
Bank pulse phase residuals are dense and smooth enough that an interpolation 
onto a uniform time series is possible. A windowed power spectrum is 
computed revealing a periodic or nearly periodic component with a period of 
$568 \pm 10$ days and a $1/f^3$ power-law noise component in pulse phase with 
a noise strength 
$S_{\phi} = (1.24 \pm 0.067) \times 10^{-16}$  $\rm cycles^2/sec^2$ 
over the analysis frequency range $f = 0.003 - 0.1$ cycles/day. 
This result deviates from past analyses which characterized the pulse phase 
timing residuals as either $1/f^4$ power-law noise or a quasiperiodic process.
The analysis was checked using the Deeter polynomial method of power spectrum
estimation that was developed for the case of nonuniform sampling, but has 
lower spectral resolution. The timing noise is consistent with a 
torque noise spectrum rising with analysis frequency as $f$ implying {\it blue} torque noise, a result not predicted by current models of pulsar 
timing noise. If the periodic or nearly periodic component is due to a binary 
companion, we find a mass function 
$f(M) = (6.8 \pm 2.4) \times 10^{-16}$ $\Msun$ and a companion mass, 
$M_c \geq 3.2 \Mearth$ assuming a Crab pulsar mass of $1.4 \Msun$.

\end{abstract}


\begin{center}
{Submitted to MNRAS 10/30/2002}
\end{center}

\section{Introduction}

Isolated rotation-powered pulsars generally exhibit a pulse frequency that
can be measured to high precision and that decreases, for the most
part smoothly, on long time scales, indicating a steady loss of angular 
momentum from the rotating neutron star. 
The deterministic spindown trend of the pulsars can be modeled by a low order 
polynomial in the pulse phase.
Apart from well known ``glitches'' in rotation rate occasionally observed in 
a few pulsars, many pulsars show fluctuations in the pulse phase about the 
deterministic trend. These rotation fluctuations or ``timing noise'', often 
do not have the form of white noise but rather exhibit a ``wandering'' 
behavior with a roughly polynomial appearance. This is not a sign of unmodeled 
higher order polynomial terms in the deterministic spindown trend, however, 
since adding more polynomial terms does not improve one's ability to predict 
the pulse phase. The nature of these rotation fluctuations is of interest 
since they provide direct observational information on the forces acting upon 
the neutron star crust. The presence of timing noise also complicates the 
detection of perturbations to the pulse phase caused by orbiting neutron star 
companions and has the potential to be mistaken for such perturbations if not 
properly understood.

The Crab pulsar spins at a rotation frequency of $30$ Hz and is the 
most well observed of all isolated pulsars. 
The Crab pulsar was first discovered in 1968 and has been subjected to 
many long term observation programs at optical and radio wavelengths
(Groth 1975a; Gullahorn et al. 1977; Lohsen 1981; Lyne, Pritchard \& Smith
1988, 1993).  
There have been several prior analyses of the long term properties of the 
pulse phase of the Crab pulsar. They all agree that the intrinsic pulse 
phase can be divided into three components: (1) A long term spin down 
that can be characterized by a low order polynomial (2) several glitches 
and (3) timing noise. 

The nature of the Crab timing noise has been investigated a 
number of times. The timing noise is revealed as the dominant component in 
the timing residuals after removal of a spindown polynomial and glitches from 
the pulse phase. The timing residuals from optical observations were compared 
to noise models consisting of random walks in pulse phase, pulse frequency 
or pulse frequency derivative (Boynton et al. 1972; Groth 1975b,c; Cordes 1980)
and found to be most consistent with a random walk in pulse frequency in all 
cases. A low resolution power spectrum was calculated for the pulse frequency 
derivative using a polynomial estimator method by Deeter (1981) (also 
displayed in Boynton 1981; 
Alpar, Nandkumar \& Pines 1986), who found a white spectrum consistent with
a random walk in pulse frequency. For a review of this work see 
D'Alessandro (1997) and Deshpande et al. (1996). In contrast, using a 
five year span of radio observations from Jodrell Bank in the 1980's, 
Lyne, Pritchard \& Smith (1988) claimed that the timing residuals were
consistent with a physically real quasiperiodic process with a period of 
$\sim 20$ months since the period did not scale with the length of 
the data being fit, as occurs with the apparent periods observed in
the residuals of polynomial fits to red noise. The disagreement between these 
various analyses has 
motivated this study of the timing properties of the Crab pulsar and a search 
for analysis techniques to better characterize the statistical nature of the 
timing noise.

We focus on power spectral analysis methods in this paper to characterize the
Crab pulsar timing residuals. The power spectral analysis is not 
straightforward due to a steep rise in power at low analysis frequencies and 
due to the nonuniform sampling of the data sets. The Crab timing residuals 
most likely contain some type of red power-law noise process, of which the 
random walk is a well-known example. Random walks have power spectra that rise 
as $1/f^2$ toward low analysis frequencies ($f$). We will refer to $f$ as 
{\it analysis} frequency to prevent confusion with the {\it pulse} 
frequency $\nu$. In general, a steep rise in 
power at low analysis frequencies can cause severe power ``leakage'' into the 
low frequency side-lobes of the frequency response of power spectral 
estimators, such as the squared amplitudes of a Fourier transform, that can
be comparable to or greater than the response at the nominal frequency of the 
estimator itself. The leaked power greatly distorts the computed power 
spectrum preventing detection of its true form unless special precautions are 
taken to control this leakage.

The nonuniformity of sampling has been severe in the pulse timing data sets 
obtained from optical observation programs that were carried out during the 
interval from 1969 to 1980. The optical observations are subject to seasonal 
and monthly interruptions due to conjunctions of the Crab pulsar with the sun 
and moon. The total length of a data set that can be analyzed using a single 
pulse phase ephemeris has been limited as well due to the occurence of 
occasional large glitches that hinder correct extrapolation of the pulse phase 
across the glitch. Long term radio monitoring of the Crab pulsar at Jodrell 
Bank from 1982 to the present has produced a set of pulse arrival times with 
much smaller gaps than in the optical data and generally has a higher sampling 
rate. In this paper we concentrate on a subset of the Jodrell Bank data set 
over the 7.5 year interval from February 1982 to August 1989 that is 
bounded by sparse data sampling prior to 1982 and a large glitch that 
occurred in 1989.

Mathematical characterization of the timing noise in the Crab and other
pulsars has frequently been in terms of red power-law noise processes
that consist of random walks in the pulse phase, pulse frequency or pulse
frequency derivative or mixtures thereof (e.g Cordes and Downs 1985). One of 
the major attractions of these models is that discrete random walk 
realizations (and summations thereof) can be easily generated numerically 
using time domain methods.  Random walks and their repeated summations (or 
integrals in the case of continuous sampling) all have ensemble averaged 
power spectra of the form $1/f^m$ where the power-law spectral index $m$ is 
an even integer. These noise models have been important for checking and 
calibrating various red noise analysis methods. We show that these time domain 
methods can be generalized to produce red-power law noise processes with 
arbitrary spectral indices (see appendix A). In addition, red (and blue) power 
law noise realizations with arbitrary spectral indices can also be generated 
using frequency domain methods. We use both time domain and frequency domain 
simulated noise realizations to check the accuracy, efficacy and calibration 
of the analysis procedures used in this study.

Analysis of pulse arrival times necessarily requires the fitting of
a low order polynomial to obtain timing residuals. When the timing
noise has a steep red power spectrum, the polynomial fitting removes 
noise power at the lowest frequencies, creating a low frequency turnover 
or peak in the power spectrum. In the time domain, this tends to produce 
residuals with a quasiperiodic appearance. The appearance of a low frequency 
quasiperiodic looking residue can easily be mistaken for a real 
quasiperiodicity or even a real periodicity by the unwary investigator when 
in fact the quasiperiodicity is purely an artifact of the polynomial 
fitting and the quasiperiod will scale with the length of the 
data fit. To detect a true low-frequency quasiperiodicity in the presence of 
detrended red noise requires more careful analysis than simple visual 
inspection of the timing residuals.    
We show that in the Crab pulsar, a real periodicity or quasiperiodicity is 
present in the power spectrum of the Crab timing residuals produced from the 
Jodrell Bank observations in addition to a red noise component.            

The analysis of Deeter (1981) involved developing and applying a method of 
computing low frequency resolution power spectra that deals with the 
problems of nonuniform sampling and red noise leakage. The method uses
polynomial fits on a hierarchy of time scales to generate a power spectrum
at octave frequency resolution. The low frequency resolution of this method 
means that any significant narrow features in the power spectrum will be unresolved and allows 
significant room for misinterpretation of the power spectral shape. We show 
that high resolution power spectra can be computed for the case of uniform 
data sampling by use of a simple windowing procedure in conjunction with
polynomial detrending to alleviate the leakage problem. The smoothness and 
dense sampling of the Crab timing residuals
produced from the Jodrell Bank observations makes possible an interpolation 
onto a uniform grid and computation of a high resolution power spectrum. This 
new analysis shows that the pulse phase timing noise is composed of two 
components: a red noise process with an approximately $1/f^3$ power density 
spectrum and a long term periodic or nearly periodic oscillation with a 
period of $568 \pm 10$ days. We confirm this result, at lower spectral 
resolution, using the Deeter polynomial method.                            


\section{The major components of the pulse phase of the Crab pulsar}  


The pulse phase of the Crab pulsar is obtained via the application
of a pulse phase ephemeris to a set of measured pulse arrival times that 
have been corrected to the solar system barycenter and corrected for other
timing delays such as those caused by interstellar dispersion of radio
waves (see Lyne, Pritchard \& Smith, 1988). For the Crab pulsar, the 
intrinsic pulse phase can be decomposed conceptually into three components: 
1) the pulsar spin frequency and its deterministic decrease, 2) glitches 
and 3) rotation fluctuations about the deterministic trend. These appear to 
be essentially independent components although the presence of glitches may be
effecting the long term slow-down process
(Lyne, Pritchard \& Smith, 1993). In addition, we will consider the possible 
presence of a longterm quasiperiodic or periodic variation in the pulse phase. 
The pulse phase can be represented by a model: 
\begin{equation}
\phi (t) = 
\phi_{S} (t) + \phi_{G} (t) + \phi_{TN} (t) + \phi_{QP} (t) + \phi_{M} (t)
\end{equation}
representing respectively the contributions of the rotation frequency and its
deterministic spindown ($\phi_{S}$), glitches ($\phi_{G}$), intrinsic 
timing noise ($\phi_{TN}$), a quasiperiodic process ($\phi_{QP}$) and 
measurement errors ($\phi_{M}$). In order to study the properties or detect the 
presence of any random or longterm periodic processes in the pulse phase we  
have to investigate the properties of the pulse phase residuals
$\phi_{res} (t) = \phi (t) - \phi_{eph} (t)$ on a time span $T$ rather than 
$\phi (t)$ directly. The pulse phase ephemeris usually consists only of a low 
order polynomial to model $\phi_{S}$, although ``glitch functions'' are 
sometimes added to fit the glitches. For the data sets analyzed in this paper, 
glitches will either be absent or only a minor perturbation on the pulse phase 
residuals that remains after subtraction of the pulse phase ephemeris.

%

The deterministic spindown of the Crab pulsar is modeled using a simple 
spindown model, $ \dot \nu =-K \nu^n, $
where $\nu $ is the rotation frequency and $n$ is the braking index. The
value of the braking index depends on the mechanism slowing down the
pulsar. For braking purely by the emission of magnetic dipole
radiation $n=3$ and for braking only by emission of a wind $n=1$. The braking 
index can be determined using
$n= {{\nu_0 \ddot \nu_0} \over {\dot \nu_0^2}}$ if $\ddot \nu_0$ can be
measured. The analytic solution to the spindown model is given by: 
\begin{equation}
\phi_{s} (t-t_0)=\dot \nu_0^{-1}\nu_0^2(2-n)^{-1}\left(
\left[1+\dot \nu_0\nu_0^{-1}(1-n)(t-t_0\right)\right]
^{ {(2-n)} \over {(1-n)}}-1)+\phi_0.
\end{equation}
The term \ $\dot \nu_0\nu_0^{-1}$\ is quite small for all known pulsars
so the analytic solution can be expanded using the binomial theorem
into a ``spindown polynomial'': 
\begin{equation}
\phi_{s} (t-t_0)=\phi_0+\nu_0(t-t_0)+{1 \over 2}\dot \nu_0
(t-t_0)^2+{1 \over 6}\ddot \nu_0(t-t_0)^3 + 
{1 \over 24}\stackrel{\ldots}{\nu_0} (t-t_0)^4 \dots.
\end{equation}
The Crab pulsar has a measured pulse frequency derivative of 
$\dot \nu_0 = - 3.862 \times10^{-10}$ Hz/s and a braking index of $2.5$ 
(Lyne, Pritchard \& Smith 1993) thus successive coefficients in the Crab
spindown polynomial will decrease in magnitude by a factor of 
$\approx 10^{-10}.$\ This implies that, if the rotation phase consisted only 
of this spindown polynomial, a polynomial fit of order $l$ will always produce 
polynomial residuals of order $l+1$, unless masked by measurement errors. 
The timing noise in the Crab pulsar was initially revealed by the presence
of polynomial-like residuals that were much larger those expected from a 
pure spindown polynomial (Boynton et al. 1972). 

\section{Observations and computation of pulse timing residuals}

A set of radio pulse arrival times for the Crab pulsar corrected to the
solar-system barycenter were obtained from Jodrell Bank (Lyne, private
communication). The data consist of 7177 arrival times collected over a nine 
year span from January 27, 1982 to January 23, 1991 (Truncated Julian Days 
4996 to 8279; TJD = JD -- 2440000.5) at five radio frequencies 
(408, 610, 930, 1420 and 1667 Mhz) with approximately $85\%$ of the arrival
times at 610 Mhz. The typical arrival time error was 9 microseconds.
These data or portions thereof have been previously presented in
Lyne and Pritchard (1987), Lyne, Pritchard \& Smith (1988) and Lyne, Smith
\& Pritchard (1992) and Lyne, Pritchard \& Smith (1993). Two glitches
are present in the data occuring on TJD 6664 with a 
$\Delta \nu / \nu \sim 10^{-10}$ (Lyne and Pritchard 1987) and
TJD 7767 (Lyne, Smith and Pritchard 1992). The second glitch is one of the
largest reported for the Crab pulsar ($\Delta \nu / \nu \sim 10^{-9}$) and the
post glitch arrival times are sufficiently distorted from their pre-glitch
behavior that only data prior to this glitch were included in the subsequent
analysis (4530 arrival times). In contrast, the glitch occuring on TJD 6664 
was a relatively small event and is discussed in section 8.

A quartic polynomial phase model of the form given by equation 3 was removed 
from the arrival times using values given in table 1 and the method described
on page 104 of Manchester and Taylor (1977). 
The errors in the parameters were determined by the method of fitting a 
quartic polynomial + sinusoid to the phase residuals determined from an 
initial phase ephemeris. See section 8 for a description of this process.

The power-law spindown model 
given by equation 2 predicts a value for $\stackrel{\ldots}{\nu_0}$ of
\begin{eqnarray}
\stackrel{\ldots}{\nu_0} & = & n(2n-1)\dot \nu_0^3/\nu_0^2 \nonumber \\
& = & -6.2 \times 10^{-31}\ {\rm Hz\ s^{-3}}
\end{eqnarray}
which is approximately {\it half} the value in table 1. The phase
contribution of the measured $\stackrel{\ldots}{\nu_0}$ term is $129$
cycles over a period of 2500 days, whereas the value predicted by
the power-law spindown model would contribute $56$ cycles. Hence 
$\stackrel{\ldots}{\nu_0}$ should be easily measureable without
much distortion caused by red noise interference. The secular spin-down
thus appears to be somewhat more complex than that predicted by a simple 
power-law spin-down model. This may be an indication of the longterm effects
on the spin down process caused by glitches (Lyne, Pritchard \& Smith (1993)).    

The arrival time data recieved from Jodrell Bank had been corrected only for 
a single value of the dispersion measure (DM) over the entire data span. 
Therefore, prior to removal of the quartic polynomial phase model, the arrival 
times were corrected using the difference between monthly DM measurements 
taken from the Crab pulsar monthly ephemeris available at the world wide 
website: http://www.jb.man.ac.uk/~pulsar/crab.html and a single DM
value of 56.743. The changes in the DM value from month-to-month caused small
discontinuities in the arrival times at the bin boundaries. This problem was 
dealt with by making a cubic spline interpolation to the set of monthly DM 
values at a higher time resolution and then computing DM values with a linear 
interpolation at the actual data times using the nearest pair of DM values
from the cubic spline interpolation. 
The resulting timing residuals are 
quite smooth and very similar to that displayed in figure 6 of Lyne, 
Pritchard \& Smith (1993). A fair number of points (130) deviated 
significantly from the local trend in the timing residuals. These were judged 
to be outliers and removed. We show the timing residuals in the top panel of
figure 1. The timing residuals are relatively smooth and display an apparent 
quasiperiodic process. The number of zero-crossings is ten, five more than 
expected if a pure spindown polynomial were present.

\section{Red Power-law noise processes}  

The timing noise in the Crab pulsar has been attributed to a random walk
in pulse frequency by several previous analyses 
(Boynton et al. 1972; Groth 1975b,c; Cordes 1980). A random walk is a 
member of the class of power-law noise processes and has an 
$1/f^2$ power-law power spectrum. In pulse phase, that is the integrated 
pulse frequency, the power spectrum steepens to a $1/f^4$ power-law. We will 
refer to these type of noise models as red power-law noise (PLN). The 
{\it a priori} knowledge that red PLN is probably present is important since 
analysis methods and interpretations that usually assume only white noise is 
present would produce erroneous results. 

Random walks in pulse phase, pulse frequency and pulse frequency 
derivative\footnote{Groth (1975b) refers to these respectively as ``phase 
noise'' (PN), ``frequency noise'' (FN) and ``slowing-down noise'' (SN), 
whereas Deeter (1984) refers to $1/f^m$ power-law noise with $m = 2, 4, 6$ as 
1st, 2nd and 3rd order red noise respectively.} 
have been considerably investigated with respect to pulsar
timing noise since these are simple models with which noise realizations
can easily be generated. They are members of the class of red PLN processes 
with $1/f^m$ spectra where $m = 2, 4, 6, \ldots$ that can be generated by 
repeated integration of white noise.
Integration of white noise processes will produce random walks which have an 
ensemble averaged power density spectrum with a power-law spectral index of 
$m=2$. Integration of the random walks will produce noise realizations with an 
ensembled averaged power-law spectral index of $m=4$. 
The time series becomes smoother with each integration and hence ``redder'' in 
appearance as the low frequency components of the time series become more
dominant. For example, a white noise process in the 
pulse frequency derivative will produce a random walk in pulse frequency and 
an integrated random walk in pulse phase.   
In general, each integration will increase the 
power-law index of the power density spectrum of the noise process by 
$2$. To see this, consider a zero mean continuous\footnote{Continuous in the
random variable sense, i.e. defined for all times $t$} white noise process,
$\epsilon (t)$, with an autocovariance function  
$< \epsilon(t)\epsilon(t+\tau) >=\sigma_{\epsilon}^{2}\delta(\tau)$,
where $\delta(t)$ is the Dirac delta-function. 
The quantity $S=\sigma_{\epsilon}^2$ is called the ``noise
strength'' and is a fixed parameter that characterizes the white noise.
The ensemble averaged power density per unit analysis frequency of 
$\epsilon (t)$ will be a constant, $<\vert F_{\epsilon}(f) \vert^2>=S$ 
where $f$ is analysis frequency and $F_{\epsilon}(f)$ is the Fourier 
transform of $\epsilon (t)$. 
The derivative theorem for Fourier transforms states that: if 
$f(t)$ has a Fourier transform $F(f)$ then $f^{'}(t)$ has a Fourier transform 
$i2\pi fF(f)$. 
Let us consider a random walk defined by:
\begin{equation} 
r_{2}(t)=\int_{-\infty}^{t} \epsilon(t^{'}) dt^{'} 
\end{equation}
where the subscript on $r(t)$ will refer to the 
power-law spectral index of the process which has a $1/f^2$ ensemble-averaged 
spectrum and the Fourier transform of $r_{2}(t)$ is 
$F_{r_{2}}(f)$. Since $\epsilon(t)$ is a continuous random process existing 
at all t on $-\infty < t < +\infty$, and $F_{\epsilon} (f)$ exists in the 
limit of the Fourier Transform of band-limited white noise, the fact that 
$r^{'}_{2}(t)=\epsilon (t)$ implies that the Fourier transform of 
$r^{'}_{2}$ is $i2\pi fF_{r_2}(f)$ and is equal to $F_{\epsilon}(f)$. 
Therefore, $F_{r_2}(f)=F_{\epsilon}(f)/{i2\pi f}$. The ensemble averaged power 
density spectrum of $r_{2}(t)$ is defined by: 
$P_{r_2}(f)=P_{\epsilon}(f)/{4\pi^2 f^2}=S/{4\pi^2 f^2}$. This argument can be 
repeated for each successive integration. Thus each integration will steepen 
the power density spectrum by a factor $1/{(4\pi^2 f^2)}$. 

Previous timing analysis of the Crab pulsar 
(Boynton et al. 1972; Groth 1975b,c; Cordes 1980 and Deeter 1981) focused on 
power-law noise processess that could be generated by the integration or 
summation of white noise realizations and so was biased toward power-law noise 
processes with even integer spectral indices. There is no {\it a priori} 
reason that the timing noise should be restricted to an even integer red 
power-law noise process and a truly general analysis method should be able to 
tell the spectral index of any power-law noise directly rather than making a 
best fit from among a set of expected indices as has been done in several of 
these past analyses. The Deeter polynomial method (Deeter \& Boynton 1982, 
Deeter 1984) can in principle determine the power-law index directly but has 
such low spectral resolution that the presence of any discrete features such 
as those due to a quasiperiodic process will produce an erroneous 
interpretation. A need exists for higher spectral resolution methods that can 
directly determine the power spectral index and distinguish the presence of 
narrow features, which we explore below.

In order to test possible analysis methods, we develop a time domain method 
to generate red power-law noise realizations with an arbitrary spectral index 
in Appendix A, which is a generalization of the discussion given by Groth 
(1975b). A useful, but unnormalized, frequency domain method which weights 
Fourier amplitudes by a power-law and then inverse Fourier transforms the 
result into the time domain has been presented by Timmer \& K\"onig (1995) 
which also generates PLN realizations with an arbitrary spectral index and is 
very straightforward to implement (see Appendix B for a normalized method). 

\section{Power Spectral Density Estimation of Red Noise Processes}  

The estimation of the power density spectrum (PDS) of a red PLN
realization presents special problems due to the steep power rise in the
spectrum at low frequencies.
The problem arises due to leakage of power in the low-frequency sidelobe of 
the frequency response of a
spectral density estimator which can greatly exceed the response at the central
or nominal frequency. Consider a noise realization
$r_m(t)$ over the time interval $\frac{-T}{2}$ to $\frac{T}{2}$. The estimated
PDS will be the squared modulus of the convolution the 
Fourier transform of the ``window function'', $W(f)$ with the intrinsic Fourier 
transform of the noise process $F(f)$:
\begin{equation}                                        
F_w(f)=\int_{-\infty}^{+\infty} F(f^{'})W(f-f^{'}) df^{'}
\end{equation}
If the time interval extended
over infinity then the Fourier transform of the window function would be a 
$\delta$-function and the intrinsic power spectrum would be recovered exactly. 
The ordinary rectangular window function:
\begin{displaymath}
w(t)=
\left\{ \begin{array}{ll} 1 & \mbox{for $|t| \leq T/2$} \\
0 & \mbox{otherwise} 
\end{array} 
\right \} 
\end{displaymath}
has a Fourier transform given by:
\begin{displaymath}
W(f)=\frac{\sin (\pi fT)}{\pi f}. 
\end{displaymath}
When $F(f)$ has a power-law form: $\frac{C}{(2\pi f)^{m}}$, the windowed
Fourier transform is given by:
\begin{equation}
F_w(f)=\int_{-\infty}^{+\infty} \frac{C}{(\pi f^{'})^{m}}
\frac{\sin \left[\pi(f-f^{'})T\right]}{2\pi (f-f^{'})} df^{'} 
\end{equation}
The power leakage problem occurs due to the singularity in the power-law noise 
spectrum at zero analysis frequency. 

In practice, the rise in power towards 
zero frequency will exhibit a cut-off due to the use of a finite, rather than 
infinite, stretch of data. The main leakage problem is caused by the large,
but finite power, at the lowest frequencies actually present. When 
$m \geq 2$, i.e. for red PLN as red or more red than a random walk, each 
frequency estimate will be dominated by the response to the power at the 
lowest frequencies. The estimated spectrum will then have a power-law spectral
index of $\approx 2$ regardless of the value of $m$. Detrending the data prior 
to power spectral estimation, by a polynomial fit for example, will decrease 
the power at the lowest frequencies and will help to alleviate the leakage 
problem irregardless of the type of power spectral estimator 
applied to the data with the tradeoff that the nondetrended form of the 
power spectrum cannot be measured directly. These issues will be further
explored in section 7.  

\section{Computation of a Windowed PDS in the Presence of Red Power-Law Noise}  

A PDS of a uniformly sampled red noise time-series can be 
computed with fast Fourier Transforms by using an appropriate window function 
to control the power leakage. The window function must be chosen to 
sufficiently reduce the side-lobe response of the power density estimator (at 
the expense of spectral resolution) so that leakage is not a major problem. 
For a given time-series,
$\{ x_k, t_k \} \ k=0,1, \ldots N-1$, the Hann window of the form 
$w_k=sin^{\alpha}(\pi k/N)$ with $\alpha=2$ has a frequency
response with sidelobes whose envelope falls-off with analysis frequency as 
$\approx 1/f^6$ (see Harris 1985). This is sufficient for computation of power 
density spectra for red noise with power-law indices up to $m \approx 5$ 
according to the moment condition criteria (Deeter \& Boynton 1984). 
Presumeably a window function could be designed to optimally suppress the 
side-lobe response while preserving maximum spectral resolution but this has 
not been pursued in this paper, since sine windows have proven to be 
simple and adequate in practice.

The actual method of PDS computation and normalization used
here involves several steps. After a window, $\hat{w}_k$, is chosen the window 
is normalized to a root-mean square of 1:
\begin{displaymath}
w_k=\frac{\hat{w}_k}{\sqrt{\frac{1}{N}\sum_{k=0}^{N-1}\hat{w}_k^2}} 
\end{displaymath} 
The PDS of $w_k x_k$ is computed as:
\begin{displaymath}
\vert \hat{a}_j \vert^2= \left \vert \sum_{k=0}^{N-1} w_k x_k exp(+2\pi ijk/N) 
\right \vert^2
\ \ \ \ \ j=-\frac{N}{2},\ldots \frac{N}{2}-1 
\end{displaymath}
By taking advantage of the fact that for real data  
$\vert \hat{a}_j \vert = \vert \hat{a}_{-j} \vert $, the two-sided PDS is 
converted to a one-sided PDS, $\vert {a}_j \vert^2$, by doubling all powers 
except $\vert \hat{a}_0 \vert^2$ and the Nyquist term 
($\vert \hat{a}_{-\frac{N}{2}}\vert^2$) when $N$ is even and eliminating all 
redundant negative frequency terms. 

A normalization factor of $\frac{1}{2N}$ is applied to the $\vert a_j \vert^2$.
This is done so that if $x_k$ is a white noise process then the mean power 
level will equal the white noise variance. For example, if $x_k$ is a white 
noise process with $<x_k^2>=\sigma_{\epsilon}^2$ and $<x_k>=0$ then:
\begin{eqnarray}
\left \langle \sum_{k=0}^{N-1} \vert w_k x_k \vert^2 \right \rangle & = &
\sum_{k=0}^{N-1} w_k^2  \langle \vert x_k \vert^2 \rangle \nonumber \\
          & = &  \sigma_\epsilon^2 \sum_{k=0}^{N-1} w_k^2 \nonumber \\
          & = &                       N \sigma_\epsilon^2 \\
\end{eqnarray}
Using Parseval's relation with the constant term $(j=0)$ deleted we have:
\begin{eqnarray}
N \sigma_\epsilon^2 & = & 
\frac{1}{N} \sum_{j=1}^{N/2} \vert a_j \vert^2 \nonumber \\
& = & \frac{1}{2}\bar{P} \\
\end{eqnarray}
where $\vert a_j \vert^2$ are the one-sided powers defined above and 
$\bar{P}$ is the mean power level. We therefore define a normalized PDS as:
\begin{equation}
P_j =\frac{1}{2N} \vert a_j \vert^2 \ \ \ \ \ j=1,\ldots \frac{N}{2} 
\end{equation} 
so the mean power level will approximately equal the white noise variance
$\sigma_\epsilon^2$ if $x_k$ is a zero-mean white noise variable.

The power-law index of the red noise noise process can be determined by
performing a linear fit to $\log P_j$ vs. $\log f_j$ where the 
$f_j=\frac{j}{N}$ are the analysis frequencies. An appropriate frequency
range must be chosen for the fit to avoid frequency regions where the
power spectrum contains substantial contributions from any other processes
present in addition to a pure power-law noise process. In this paper
we used linear least-squares fitting to estimate the slope. The data points 
$\log P_j$ do not have a gaussian distribution about a mean value 
$< \log P_j >$ as assumed in ordinary least-squares fitting, so some 
distortion in the slope and slope error 
estimates will be introduced. The actual distribution of the data points about 
a mean value is similar to gaussian but possesses a low value tail that 
results from taking the log of a chi-squared distribution with two degrees of 
freedom. The actual slope estimation procedure used here involved an initial 
fit with assumed unit errors on $\log P$, a calculation of the deviation 
of the fit residuals and then a new fit with uniform errors equal to the 
measured 
deviation. Monte Carlo simulations of this procedure on simulated data sets 
of 10000 sample realizations each for a variety of input power-law indices 
showed that no bias existed in the estimated mean slope value but the 
slope errors had been overestimated by a factor of 1.7. Hence, the slope 
errors quoted in this paper have been divided by 1.7.

The power density spectrum can be converted to a flattened form to allow
clearer study of features unrelated to the power-law form. We chose a
normalization that allows the noise strength to be estimated directly from
the mean power-level over a suitable analysis frequency range. Given a 
power-law index $m$ the power spectrum is multiplied by 
$(2 \pi f_j)^m$. The analysis frequency is converted to physical units as 
$\nu_j = \left (\frac{f_j}{\Delta t}\right )$ where $\Delta t=t_{n+1}-t_n$. 
The flattened PDS is then normalized so the mean power level approximately 
equals the noise strength in physical units by multiplying by a factor
$(\frac{1}{\Delta t})^{m-1}.$ This is a
combination of a factor of $(\frac{1}{\Delta t})^{m}$ to flatten the spectrum,
a factor of $(\frac{1}{\Delta t})^{-2}$ to correct the power via the similarity
theorem for the rescaling of the frequency and a factor of 
$(\frac{1}{\Delta t})$ to make the mean power level equal the noise strength 
in the presence of pure power-law noise. The extra factor of $(\frac{1}{\Delta
t})$ arises from the definition of the noise strength $S$ (see section 7 and
Appendix A). The total set of
operations on the one-sided PDS can be combined as:
\begin{equation}
P_{j,flattened} = 
\left (\frac{1}{2N}\right ) \left (2 \pi f_j\right )^m \left (
\frac{1}{\Delta t}\right )^{m-1}
\vert a_j \vert^2 \ \ \ \ j=1, \ldots \frac{N}{2}-1. 
\end{equation}
Note that the power-law spectral index $m$ is a free parameter supplied by 
the user. The units of the noise strength are given by the units of $a_j$ 
and $\Delta t$. For example, a random walk in pulse phase (measured in
pulse cycles with time measured in seconds) will have $m=2$ and hence
a noise strength $S = \left < P_{flattened} \right >$ with the physical units 
of $\rm cycles^2/s$, whereas a random walk in pulse frequency (m=4) will
have the physical units of $\rm cycles^2/s^3$. 

\section{Power Spectrum Estimation of Simulated Red Noise Realizations
with Uniform Sampling}






We explored the practical estimation of power spectra of red 
PLN processes using simulated PLN time-series. This 
problem involved determining methods to generate PLN processes and
then seeing if the predicted spectra and noise strength could be correctly
recovered. Neither the generation of red PLN processes 
with non-even integer spectral indices nor high resolution estimation of
their power spectra has been previously well developed, so sorting out the 
presence of possible systematic effects and attributing them to either the 
noise generation method or the spectral estimation method required extensive 
testing.

We implemented the time domain method of generating discrete red PLN
realizations described in Appendix A. With this method a band limited
white noise realization $\epsilon(t)$ is used as a basis upon which a 
power-law weight function acts to generate the red noise realization. The 
white noise is a zero mean process with a gaussian distribution of impulse 
amplitudes generated at times $t_k$ with nonuniform time steps 
$(\Delta t = t_{k+1} - t_k)$, where $\Delta t$ is drawn from the exponential 
waiting time distribution of a Poisson process. The chosen step rate $R_g$ and 
variance of the step amplitude distribution $\sigma^2_{\epsilon}$ determine 
the noise strength $S=R_g \sigma^2_{\epsilon}$. The power-law weight function
then acts upon the discrete white noise realization to produce a set of
accumulated steps that can be sampled at an arbitrary set of times $t_i$ that 
are independent of the times of the white noise impulses. For the case of a 
random walk, one will have a set of rectangular steps of varying amplitude
and duration, whereas for $1/f^3$ noise, the individual steps will have a form 
$\propto t^{\frac{1}{2}}$. The appearance of the simulated noise realization 
for the same noise strength and power-law index varies appreciably depending 
on the sampling and step rates. For example, sampling at a rate greater than 
the white noise step rate $R_g$ allows individual steps to be resolved. In
figure 2 we portray a set of discrete red power law noise samples with resolved
steps.  

In figure 3 we show a set of red noise realizations generated from the same 
underlying white noise realization. As the power-spectrum becomes ``redder'' 
the time series takes on an increasingly smooth appearance that arises from the 
increasing dominance of the lowest frequency components. The right-hand set 
of panels shows the same noise realizations after cubic polynomial detrending. 
A quasiperiodicity is apparent in the residuals in some cases that is a
result of the polynomial detrending. Polynomial fits on shorter or longer 
timescales produce similar looking residuals and hence a ``pseudoperiodicity''
that scales with the length of the red noise time-series.                   

In the top panel of figure 4 we show an example of a $1/f^3$ noise realization 
that has been detrended by a cubic polynomial fit. All the time-series 
discussed in this section are uniformly sampled at a rate approximately twice 
the step rate and the results discussed apply strictly only for uniform 
sampling. In the middle panel we show the computed power spectrum (rough 
histogram) computed using a Hann window with $\alpha=2$. Lying above this is a 
smoother histogram that shows the average power spectrum of 1024 similar noise 
realizations (shifted in power by 3 decades for display purposes). The 
simulations showed that the input power spectral slope and noise strength 
could be correctly recovered. The spectrum shows a clear turnover at low 
frequencies caused by the removal of low frequency power by the cubic 
polynomial detrending and this results in residuals that appear quasiperiodic.
The average analysis frequency of the peak is purely related to the length of 
the data span fit. In the bottom panel we 
show the same power spectra after having been flattened as described in section 6 assuming a power-law index 
$m=3$. We now see that the effect of the polynomial detrending is to 
{\it decrease} low frequency power in the flattened spectrum relative to 
higher frequencies. A high frequency upturn in the average spectrum is also 
present that is the result of aliased power from above the Nyquist frequency.

Examination of indvidual power spectra showed that cubic detrended realizations 
with a low number of zero-crossings (4-5) often displayed a slight excess of 
low frequency power with respect to the higher frequencies whereas those with 
higher numbers of zero-crossings tended to show a power deficit. This is 
an easily understood byproduct of the goodness of the cubic polynomial fit to
the noise realization (i.e. better fits cause larger numbers of 
zero-crossings in the residuals). A second related effect is to produce a 
negative correlation of the mean squared value of the residuals with the 
number of zero-crossings, confirmed by simulations and is a general result for 
polynomial detrending of red noise. The degree of the polynomial fitting 
exhibited a weak positive correlation with the frequency of the power peak 
in the unflattened spectrum. This is expected since quartic and higher order 
polynomial detrending simply removes more low frequency power and hence
shifts the power peak frequency to higher frequencies and results
in a greater deficit of low frequency power in the flattened power spectrum. 

Several effects of windowing and polynomial detrending are illustrated
in figure 5 which displays averages of 1024 flattened power spectra for
PLN realizations of unit strength with added white noise and a high frequency
sinusoid. The power spectra were made
using a Hann window with $\alpha=2$. 
First, the polynomial detrended power spectra
show clearly the decrease in low frequency power relative to
the power spectra produced from undetrended noise realizations. The 
relative power drop increases as the power-law index increases consistent
with the fact that redder realizations are smoother and thus better fit by
polynomials. Secondly, the undetrended power spectra show an excess of power at
low frequencies which increases with the power-law index. This is probably
a power leakage effect caused by leakage into the highest sidelobes of the
power estimator which will be largest at low frequencies where most power is
concentrated. Thirdly, another leakage effect
is also readily apparent in figure 5 illustrated by the spectra of the 
undetrended realizations: namely that for power spectral estimation at
higher frequencies power leaks into the entire low frequency side of
the estimator and hence tends to get worse as the estimator's nominal
frequency increases. This effect eventually causes the spectral slope to 
change from the predicted power-law to a much lower value $(\approx 2)$
when the increasing steepness of the power density spectrum causes the leaked 
power to exceed the rate of sidelobe fall-off of the power spectral estimator 
which for the Hann window with ($\alpha=2$) is $\approx 1/f^6$.
The polynomial detrending, by decreasing the lowest frequency power, which
is the largest source of power leakage, decreases substantially both the local 
sidelobe leakage and the integrated leakage on the low frequency side of the 
estimator.

The high frequency area of figure 5 illustrates the dominance of the 
added white noise component at high frequencies as the red noise power level
drops below that of the mean white noise power level. The ``crossover''
frequency is that power level where the mean red noise power level equals 
that of the white noise and occurs in this case near the frequency $f=0.2$.
The high frequency sinusoid added to the noise realizations
clearly shows up as the narrow peak near $f=0.1$. In general, the correct
sinusoid frequency could be recovered from the flattened spectra with no
more bias than in ordinary power spectral estimation. 

We also tried other windowing functions such as the Hann window with
$\alpha=1$. This window has better frequency resolution but worse
leakage protection than for $\alpha=2$. Major leakage effects began to be 
noticeable for power-law indices for $m>3.5$ for power spectra of nondetrended
noise realizations and for $m>4.4$ for detrended realizations. This is
expected since the sidelobes fall-off as $\approx 1/f^4$. Other
window functions such as the Hamming, Kaiser-Bessel or Bohman were examined as 
well but rejected due to their poor leakage protection. The Blackman-Harris
window offered leakage protection comparable or better than the cosine windows 
in some frequency regimes but had noticeably worse frequency resolution due
to the large width of the main lobe of the frequency response and so was
rejected as well.

A second method of generating red PLN has been proposed by 
Timmer \& K\"onig (1995) using a method that creates PLN realizations by 
generating the real and imaginary values of a set of Fourier amplitudes and 
then weighting them by a power-law. The amplitudes are then Fourier 
transformed back into the time-domain. The use of a second method of 
generating PLN was useful in helping to differentiate effects caused by the 
spectral estimation technique versus those caused by the noise generation 
method. Direct comparison between the two noise generation methods was at 
first complicated by the lack of a proper normalization for the 
Timmer \& K\"onig generated realizations. We determined a normalization for 
this method to generate noise with a predetermined noise strength consistent 
with that used for the time-domain method. See Appendix B for a more complete 
description.

The Timmer-K\"{o}nig method generates a time-series from a discrete set of 
Fourier frequencies and is therefore missing power from frequencies not within 
the discrete set. Because of this, there is no power at frequencies higher 
than the Nyquist frequency, or lower than the lowest frequency $1/T$, where 
$T$ is the total data timespan, as there would be in a more realistic segment 
of PLN. If the time-series is resampled at a higher rate, then a small 
aliasing effect can be observed in the power spectrum at the highest 
frequencies. Likewise, if the time-series is resampled at a lower rate then a 
deficit of power is apparent at the highest frequencies. If one uses the full 
time-series, then the endpoints and the derivatives of the lowest order 
frequency terms nearly match when the time series is cyclically repeated, 
which greatly reduces the power leakage problem. Thus we discarded the second 
half of the time-series to better represent true power-law noise. Without 
discarding the second half, flattened power-spectrum with the correct power 
and slope can be produced using only a rectangular window for power-law 
indices much greater than $m=2$, e.g. upto at least $m=5.5$! This is just
a result of artificially constructing red noise only from harmonics that are 
exactly periodic on the analysis interval which causes all leaked power to 
fall exactly on the zero's of the {\it sinc} function frequency response.

Average power spectra similar to that of figure 5 were calculated for
noise realizations generated by the Timmer-K\"{o}nig method. These
power spectra, similarly to those in figure 5, showed an excess in low
frequency power for undetrended realizations and a subsequent drop in the
power as a result of detrending. Major leakage effects set in at similar 
power-law indices. The only real differences were the lack of aliased high 
frequency power and different high frequency distortion effects for power-law
indices $m \geq 5.5$ where the Hann window with $\alpha =2$ ceases to provide
adequate leakage protection. 


\section{Analysis of the 1982-1989 Crab radio timing residuals}


\subsection{Windowed Power Spectrum}

We calculated a windowed power spectrum of the Jodrell Bank Crab timing 
residuals. The dense sampling and relative smoothness of the timing residual 
suggested that interpolation onto a uniform grid was possible without 
significant distortions. The timing residuals were first averaged on a one-day 
time scale, mainly to reduce the number of points in the densely sampled region
after the 1986 glitch (see Lyne, Pritchard \& Smith 1993 and figure 6, middle
panel). The data were then interpolated onto a uniform grid using a cubic 
spline interpolation with $\Delta t = 1.346$ days. Two samples comparing the 
uniformly gridded phase residuals with the unprocessed phase residuals are 
shown in figure 7. Using the method described in section 6 a power spectrum
was calculated using a Hann window with $\alpha = 1$. The uniformly
binned time-series, windowed power spectrum and flattened power spectrum
are displayed in figure 6. The power spectrum of the timing noise shows
a power-law with a spectral index of $m = 3.09 \pm 0.05$. The flattened power
spectrum shows a clear low frequency peak at a frequency $f \approx 0.0018$
cycles/day. In addition, the flattened power spectrum exhibits a decrease 
in low frequency power due to the removal of the polyonomial trend and an 
upturn at high frequencies due to white noise in the pulse phase. A power 
spectrum was also calculated using a Hann window with $\alpha = 2$ (see bottom 
panel of figure 1) which provides
better protection against red noise leakage but worse frequency resolution.
Essentially the same power spectrum was produced but with worse resolution
of the low frequency peak.

The power spectrum of the Crab timing residuals is consistent 
with the presence of a power-law noise component with a spectral index 
$m \approx 3$. The noise strength is given by the mean white noise level in 
the flattened spectrum. The averaged power in the flattened spectrum over the 
frequency range $0.003 - 0.1$ cycles/day is $(9.26 \pm 0.5) \times 10^{-7}$
$\rm cycles^2/day^2$. This implies a noise strength 
$S = (1.24 \pm 0.07) \times 10^{-16}$ $\rm cycles^2/sec^2$. The timing noise 
is consistent with a torque noise spectrum rising with analysis frequency as 
$\approx f$, implying {\it blue} torque noise, rather than white torque noise 
as found  by Deeter (1981), Cordes (1980) and Groth (1975c). For the numerical 
1st derivative of the uniformly binned time-series we repeated the power 
spectral analysis described above and found a spectral index 
$m = 1.12 \pm 0.05$  
and a mean flattened noise power (assuming $m=1$) of 
$(8.96 \pm 0.5) \times 10^{-7}$ $\rm Hz^2$. Likewise, the numerical 2nd 
derivative had a spectral index $m=-0.86 \pm 0.05$  
with a mean flattened noise power (assuming $m=-1$) of 
$(8.77 \pm 0.5) \times 10^{-7}$ $\rm (Hz/s)^2 day^2$. Both these results
are consistent with that for the pulse phase. A large peak in the 
power showed up clearly at the same analysis frequency in the flattened 
spectra for all three cases. The observed power spectrum of the pulse phase 
residuals is consistent with the presence of a power-law noise process, but we 
point out that this does not mean a power-law noise process such as those 
described in appendix A is the correct form for the noise since there
may be other noise processes that could produce similar power spectra.

The cubic spline interpolation onto a uniform grid could introduce some 
distortions in the computed power-spectrum, in spite of the relative 
smoothness of the Crab timing residuals. We checked the correctness of the 
analysis procedure described above using Monte Carlo simulations. A noise 
sample was generated by using the times of the Crab timing residuals after 
outliers had been removed. Using these sample times we generated a realization 
of $1/f^3$ power-law noise using the procedure described in Appendix A with a 
noise strength $S=1.0 \times 10^{-6}$ $\rm cycles^2/day^2$ and a step rate 
$R_g = 0.7434$ $\rm day^{-1}$. To this we added a white noise realization with a 
variance equal to the variance of the Crab timing measurement noise and a 
cosinusoid with parameters given in table 2. The realization was then fit by 
a quartic polynomial, averaged on a one day timescale and interpolated onto 
a uniformly gridded time-series using a cubic spline following the same
procedure used in analysis of the actual Crab timing data. Power spectra 
of the resulting time-series were computed using Hann windows with 
$\alpha =2$.   

We generated a sample run of 512 simulated noise realizations, computed power 
spectra, and measured the power spectrum slope and noise power level in the 
flattened power spectrum over the frequency range $0.003$ to $0.1$ cycles/day.
This frequency range avoided the low frequency power contribution of 
the cosinusoid and the high frequency white noise contribution. We found a 
mean PDS slope of $-3.03 \pm 0.15$ 
and a mean power level of 
$(0.97 \pm 0.1) \times 10^{-6}$ $\rm cycles^2/day^2$, consistent with the 
input values. The peak frequency of the input cosinusoid was correctly 
recovered. 
A second similar run of 512 sample noise realizations without the low
frequency cosinusoid was created and analyzed. A mean PDS slope of 
$-3.03 \pm 0.14$ and a mean power level of 
$(0.98 \pm 0.1) \times 10^{-6}$ $\rm cycles^2/day^2$ was found.
Any distortions due to aliasing would show up in the high
frequency portion of the spectrum where the effects of white measurement 
noise tend to dominate and hence this is not a useful portion of the spectrum
in any case. We therefore conclude that no significant distortions to the power 
spectra were introduced by the data analysis procedure described above.

\subsection{Quasiperiodicity}

The large low frequency peak in the Crab power spectrum appears to be a real
feature and not simply a result of polynomial detrending of red noise. 
Detrended red noise should show a significant {\it decrease} in the low 
frequency power in the flattened spectrum relative to higher frequencies
rather than an increase, as figure 4 demonstrates. The quartic polynomial 
removed from the Crab timing measurements will cause an even greater drop 
in low frequency power in the flattened spectrum relative to that shown in 
figure 4 where cubic detrending was used.    
To investigate the significance of the power spectral peak we assumed 
the ``white noise'' in the flattened power spectrum followed a chi-square 
distribution. To check this assumption we made a histogram of the flattened 
power over the frequency range $f = 0.003 - 0.1$. The powers did indeed 
follow an exponential distribution. The peak at $f = 0.0018$ cycles/day has 
a probability of chance occurence of less than $1 \times 10^{-6}$. 

To determine whether the power spectrum peak was due to a periodic or 
quasiperiodic process we performed a sinusoidal fit to the Crab phase
residuals in the low frequency region around the power spectrum peak.
We represented the Crab phase residuals in the time domain with a model given 
by \begin{equation}
\phi_{\rm model}(t) = \phi_0 + \nu_0 \Delta t + \sum_{j=1}^4 \frac{d^j \nu}{d
t^j} \frac{\Delta t^{j+1}}{(j+1)!} +
A \cos(2 \pi f \Delta t) + B \sin(2 \pi f \Delta t)
\end{equation}
where $f$ is the sinusoid frequency and $\Delta t = t - t_0$ where $t_0 = {\rm
TJD}\ 6390$ is an epoch.
This model was fit to the data by doing a modified-Marquardt fit to the linear
terms for each sinusoid frequency in a grid. For each frequency, a fit 
statistic was computed, given by
\begin{equation}
\Delta \chi^2 = \chi^2_{\rm poly} - \sum_{i=1}^N \frac{|\phi_i - \phi_{\rm model} 
(t_i)|^2}{\sigma_i^2}
\end{equation}
where $\chi^2_{\rm poly}$ is computed for a fourth order polynomial model 
(no sinusoid) and $\phi_i$ are the Crab phase residuals with errors $\sigma_i$.

First sinusoid frequencies in a wide range (0.5 to 4.0 cycles per 1000 days)
were searched. 
This search found a peak in $\Delta \chi^2$ at $f \approx 1.76$
cycles/kday (equivalent to a period of 568 days) and a gradual increase in 
$\Delta \chi^2$ as the frequency decreased below about 1 cycle/kday. Next a 
narrower search (1.26-2.26 cycles/kday), centered on the peak was performed to
refine the fit. The best fit with frequency was $f \approx 1.762$ cycles/kday 
(or $P \approx 567.5$ days) with a sinusoid amplitude 
$(A^2+B^2)^{1/2} \approx 0.26$ $\rm cycles$.

To estimate the probability of finding a peak at this frequency by chance
and to place errors on the parameters, two sets of Monte-Carlo simulations were
performed. In the first set of simulations, the model above was fitted to
10000 red-noise realizations and the best-fit frequencies were retained. 
The red noise consisted of simulated power-law noise using the time-domain
method described in Appendix A with a power-law index
$m=3$ and noise strength $S=1.0 \times 10^{-6}$ $\rm cycles^2/day^2$, equivalent
to that found for the Crab pulsar. The same data sampling as the Jodrell Bank 
data was used. A distribution of the best-fit frequencies was obtained that 
had a maximum near the frequency $0.88$ cycles/kday and a broad low
frequency tail. In the second set of simulations a sinusoid with the parameters
in Table 2 was included. The resulting distribution of best-fit frequencies
was tightly clustered around a maximum frequency of 
$1.76$ cycles/kday. For the case with no sinusoid present, a total of 
69 trials fell in the frequency range 1.736 to 1.790 cycles/kday, the full 
width of the distribution of best-fit frequencies with the sinusoid present, 
suggesting a probability of chance occurrence of $6.9 \times 10^{-3}$. 
However, the largest sinusoid amplitude measured in this frequency range was 
$\sim$ 0.11, about 2.3 times smaller than the amplitude measured in the actual 
Crab data. Also, the largest $\Delta \chi^2 = 6.9 \times 10^7$, was about 5.4 
times smaller than that measured in the data. These additional factors suggest 
that the probability that this peak is due to some systematic effect is 
smaller than $6.9 \times 10^{-3}$. 

The errors on the fitted sinusoid parameters were estimated from the set of
10000 fits to the simulated red-noise realizations with a sinusoid present.
The obtained parameter distributions were approximately gaussian and parameter 
errors were determined from gaussian fits to the distributions for the 
sinusoid frequency, cosine amplitude, and sine amplitude respectively. 
The parameters and their errors were also converted to their equivalent 
representation for a single cosine and phase\footnote{
$\phi_{\rm QP}(t-t_0) = A_0\cos(2\pi (t-t_0)/P - \Phi_0)$ where the single 
cosine amplitude is given by $A_0 = (A^2 + B^2)^\frac{1}{2}.$}. 
Table 2 lists the best fit parameters along with their estimated errors. 


The low frequency oscillation in the 1982-1989 Crab phase residuals appears to 
be nearly periodic. This is suggested by the fact that when the power spectrum 
is recomputed after removal of the best fit sinusoid, essentially all excess 
power in the frequency range of the peak is removed. However, since less than 
five cycles are present in the stretch of Jodrell Bank data analysed here, 
and longer data stretches are interrupted by the occurrence of data gaps or 
large glitches, the long term coherence of the low frequency oscillation is 
not known. 

\subsection{The 1986 Glitch}



We investigated the effect of the 1986 glitch (Lyne, Pritchard \& Smith 1993) 
on the power spectrum. We tried simply deleting the glitch from the phase
residuals or fitting a function to the glitch in order to smooth over the
glitch interval. We excised the glitch over the interval TJD 6664.4 to TJD
6746.0. When this was done the spline interpolation filled in the empty gap
with a linear segment. The computed window power spectrum (using a Hann
window with $\alpha=1$) exhibited a slight
change in slope from $-3.09 \pm 0.05$ to $-3.12 \pm 0.06$ over the analysis 
frequency range $0.003 - 0.1$ cycles/day. The mean power in flattened power
spectrum dropped significantly from a level of $9.26 (\pm 0.49) \times 10^{-7}$ 
to $5.3 (\pm 0.3) \times 10^{-7}$ ${\rm cycles^2/day^3}$. The power at
highest and lowest frequency portions of the PDS was unaffected. In contrast,
when random sections of data of equal duration were removed, there was
almost no effect on either the power spectral slope or the mean power level
in the flattened spectrum.   

The second method used a fit to the glitch involving several steps. First the
phase residuals from the preglitch interval one month prior to the glitch were 
fit with a linear function. This was extrapolated into the glitch interval
and subtracted. A glitch function of the form:
$$ \phi(t)=86400 \nu_1 \tau_1 (\exp (-t/\tau_1) - 1) $$ where 
$\nu_1 = 1.73 \times 10^{-7}$ $Hz$ and $\tau_1 = 6.3$ $days$ was then subtracted. A rising phase residual was left that was fitted with a quadratic
polynomial and subtracted as well, leaving flat phase residuals with no
systematic trends. When this set of functions was removed from the phase
residuals, the phase residuals exhibited a linear trend in this interval. The 
result on the computed power spectrum was essentially the same as in the case of
the simple excising above: a slope of $-3.11 \pm 0.06$ and mean power level
in the flattened power spectrum of $5.6 (\pm 0.3) \times 10^{-7}$ 
${\rm cycles^2/day^3}$ over the analysis frequency range $0.003 - 0.1$ 
cycles/day. 

Either method of treating the glitch showed that a considerable amount of
power resulted from the glitch but without any change in the power-law
slope. Thus the noise strength estimated for the timing noise may be
too high by a factor of two.

\section{Deeter Polynomial Power Spectral Estimation}

We computed power spectra of red PLN processes and of the Crab timing 
residuals using the Deeter polynomial method and compared them to power 
spectra computed using windowing methods. We did this as a cross check on 
both power spectral estimation procedures and as a direct comparison with
previously reported Crab timing analysis. The Deeter polynomial method 
of power spectrum estimation is directly applicable in the case of 
nonuniformly sampled data, unlike windowing methods, but produces power
spectra of low spectral resolution. After comparing a number of 
procedures for estimating the power spectrum of a red-noise process, 
Deeter (1984) concluded that a technique based upon
sampling with orthonormal polynomials on a hierarchy of time scales produces
a good power spectrum with modest frequency resolution.  Specifically, Deeter
(1984) proposed constructing power spectra with approximately octave frequency
resolution by iteratively partitioning a data set into increasing numbers of 
shorter segments. For each segment, an amplitude is obtained by constructing 
a set of polynomials that are orthonormal on the segment and using the highest 
order polynomial from the set to ``sample'' the data by forming the sum of the 
product of the time series with the highest order polynomial, i.e.,
\begin{equation}
a = \sum_{k}\phi_kh_k,
\end{equation}
where $\phi_k$ is the measured phase and $h_k = p(t_k)/\sigma^2_k$,
where $p(t_k)$ is the 
(highest order orthonormal) polynomial evaluated at time $t_k$, the time of
the $k$th phase, and $\sigma_k$ is the measurement uncertainty of the $k$th 
phase.
The square of the amplitude, $a$, is an estimate of
the power in the segment.  For each partitioning, the average of the powers
in the segments is an estimate of the power in the time series, at the time
scale of the segments.  By repeating the procedure for each partitioning,
and by converting time scales to frequencies, a power spectrum 
of the white noise underlying the red noise is constructed.

For a given set of times and weights, there is a unique set of orthonormal
polynomials.  For example, for the continuous interval $[-1,1]$ with uniform
weight, they are the Legendre polynomials.
It is not difficult to show that the amplitude obtained by sampling with
the highest order polynomial in a set of orthonormal polynomials is equal to
the coefficient of the highest order term in a polynomial fit to the data 
with the same order, $l$, as the highest order orthonormal polynomial.
We chose to fit, rather than sample, and refer to this technique
as the {\it Deeter polynomial power spectrum estimation method} or more simply
as the {\it Deeter polynomial method}. We have applied this technique
in the case of accretion-powered pulsars in Bildsten et al. 1997.

As a direct comparison of the Deeter polynomial method to the windowing
method we applied each method to the same simulated red PLN time-series.
In figure 8 we show the average power spectra computed using each 
technique for the same 32 uniformly sampled noise realizations. The noise
is $1/f^3$ PLN generated using the time domain method and detrended by a cubic
polynomial fit. The spectra have been flattened for the case of a $1/f^3$
power-law, which involves dividing the Deeter polynomial power spectrum
by a factor $2 \pi f$. Sinusoids have also been added in four of the five
spectra displayed. In general, when pure red PLN was present the two techniques
gave consistent results as can be seen in the third spectra from the top,
although the normalization of the Deeter polynomial method has to be
adjusted somewhat depending on the assumed noise present. 
When strong discrete features were present in the power spectra, considerable
power leakage occurred due to the low spectral resolution of the
Deeter polynomial estimators and hence distorted the power spectra. In general,
the presence of significant discrete features can cause considerable ambiguity 
in the interpretation of the Deeter polynomial power spectra. 

A Deeter polynomial power spectra was computed for the Jodrell Bank Crab
timing residuals and is displayed in figure 1. The Crab timing residuals
used in the computation are shown in the top panel of figure 1 and were 
subjected only to 
the dispersion measure corrections and removal of outliers described in
section 3. No averaging or interpolation was done. The computed power 
spectrum for the pulse frequency rate of the phases is shown in the middle 
panel of figure 1. The power contributed by measurement noise is shown by
the triangles and has not been substracted off from the total power shown
by the solid circles, which is the convention used in this paper. If the 
timing residuals consisted of pure $1/f^4$ noise in pulse phase as claimed 
by several previous analysis the power spectrum would be flat. A distinct 
rise with frequency can be observed at frequencies greater than log 
$f = -2.5$, although the large amount of power at log $f=-2.9$ might cause 
one to argue that the spectrum is actually flat. When the spectrum is divided 
by $2 \pi f$ (bottom panel of figure 1), which will produce a flat spectrum 
if $1/f^3$ noise is present, a flat spectrum is indeed produced except for a 
peak near log $f=-2.9$. For comparison, we also plot the power spectrum 
(histogram) computed using a Hann window with $\alpha=2$ for the uniformly 
sampled time series displayed in the top panel of figure 6. The power spectra 
are similar except for a deficit in power in the Deeter polynomial spectrum 
near log $f=-2.6$, relative to the windowed spectrum.

Since past analysis of the Princeton optical timing observations have 
concluded that the timing noise in the Crab pulse phase is $1/f^4$ noise
(Groth 1975c, Cordes 1980, Deeter 1981) rather than the $1/f^3$ noise found 
above, we also investigated this data set to see why this was the case. In 
figure 9b we show a Deeter polynomial power spectrum calculated using the same 
Princeton optical data set used by Deeter (1981), which is displayed in
figure 9a. The power spectrum calculated by Deeter can be found as figure 8 
in Alpar, Nandkumar \& Pines (1986). The spectrum is calculated for the
pulse frequency derivative and hence will be flat for pure $1/f^4$ noise
in pulse phase. The spectrum in our figure 9 is relatively flat but
shows an increase in power at low frequencies. When flattened under the
assumption of $1/f^3$ noise we see that the spectrum is again relatively
flat with a relatively greater excess of power at low frequencies very 
similar to that found for the Jodrell Bank data set. However, as the 
Princeton data set is somewhat shorter than the Jodrell Bank data set, 
the excess low frequency power appears more as a curvature in the 
spectrum rather than a discrete peak. We therefore conclude that the
spectrum is consistent with either an interpretation as $1/f^4$ noise in
pulse phase or as $1/f^3$ in pulse phase combined with a low frequency
periodic or quasiperiodic process. In the absence of any motivation to assume
that a large amplitude low frequency quasiperiodic process was present, 
clearly it would be more natural to assume that only $1/f^4$ noise in pulse 
phase was present, although Groth (1975c) did speculate about the presence of 
a small contribution from $1/f^2$ noise in pulse phase. 

A cubic spline interpolation of the Princeton optical timing
residuals onto a uniform time series was made and a windowed power spectrum
computed for the sake of completeness.
The interpolated time series is shown in figure 9a offset below the actual
timing measurements. When a Hann windowed power spectrum with $\alpha=2$ was 
computed, a slope of $-3.2 \pm 0.08$ between the analysis frequency range
of $0.003 - 0.1$ cycles/day. The power spectrum flattened under the assumption
of $1/f^3$ noise is shown as the histogram in figure 9c. For the same frequency
range, a mean noise
level of $(4.9 \pm 0.37) \times 10^{-7}$ $\rm cycles^2/day^2$ is measured, about
half that for the Jodrell Bank phase residuals. The windowed power spectrum 
is remarkably similar to the Deeter polynomial power spectrum and shows a 
similar large low frequency bump. The interpolated time series is clearly in 
error from the true time series since the gaps tend to be filled with linear 
segments. However, neither method of power spectrum computation has information within the gaps and the
lowest order polynomials in the Deeter polynomial power spectrum also involve 
fitting across gaps and hence also involve a form of interpolation. Since the 
data are likely to be relatively smooth at the lowest analysis frequencies, 
these interpolations may not introduce any major distortions to the power 
spectrum.

\section{Discussion}


The timing noise in the Crab pulsar has been claimed to be either a
white torque noise process or a quasiperiodic process. 
As a result of the analysis presented here, the timing noise or rotational 
irregularities in the pulse phase of the Crab pulsar appear to be composed 
of two components: a noise process with a $1/f^3$ power spectrum and a 
periodic or nearly periodic process with a period of approximately $568 \pm 10$ 
days. A noise process with a $1/f^3$ power spectrum in pulse phase becomes a
process with a spectrum rising as $f$ in the pulse frequency derivative, so 
the timing noise in the pulse frequency derivative (or torque for the case of 
a constant moment of inertia for the neutron star) has a {\it blue} power 
spectrum. 

The argument for $1/f^3$ noise rests on the windowed power spectrum of 
the timing residuals obtained from the Jodrell Bank radio observations
over the period 1982 - 1989. The windowed method of power spectrum 
computation was thoroughly checked with simulated noise data and appears
to be a quite reliable method in this situation of relatively smooth
residuals and high density sampling. The result is reinforced by a Deeter 
polynomial power spectrum of this same data which yielded a consistent 
result. A Deeter polynomial power spectral analysis of the Princeton 
Crab optical timing data previously analyzed by several investigators
(Groth 1975b,c, Cordes 1980 and Deeter 1981), showed that this shorter
and more sparsely sampled data set was consistent with either a pure $1/f^4$
noise process or a $1/f^3$ noise process with a large excess of low
frequency power, thus explaining why previous analyses chose the simpler
interpretation of pure $1/f^4$ noise.

Many theoretical studies have proposed models for the timing noise
observed in radio pulsars involving internal or external torques acting on
the neutron star crust. A good summary and references can be found in
D'Alessandro (1997). Many of these models have been developed 
to the point that they make predictions about the form of the power 
spectrum of the torque noise. In the vortex creep model (see Alpar, 
Nandkumar \& Pines (1986)), several predicted torque power spectra 
are proposed that are flat
with a transition over some characteristic frequency range to another
flat region or are flat at high frequencies but transition over to
a $1/f^2$ power-law at low frequencies. None of
these models match the spectrum observed for the Crab pulsar. In the
internal torque noise model of Jones (1990), the predicted power for the 
torque noise is a superposition of red and white components and also does 
not match the spectrum observed for the Crab pulsar. The external torque
noise model of Cheng (1987) predicts either a blue or white spectrum 
for the torque noise but the blue component is a $f^2$ power-law rather
than $f$. There does not seem to be any current model that predicts
a blue pulse frequency derivative noise power spectrum rising as $f$.

An excess of power in both the windowed and Deeter polynomial
power spectrum is apparent at low frequencies indicating the additional
presence of a strong quasiperiodic or periodic process. A known property
of red power-law noise processes (or any noise process with a red spectrum)
is that polynomial detrending will produce quasiperiodic residuals that
are purely an artifact of the detrending. These quasiperiodicities 
have occasionally been considered real by investigators unfamiliar with
this property of red noise. However, we show the quasiperiodicity present 
in Jodrell Bank Crab timing residuals is real. This is indicated by the
presence of a peak in the low frequency power in the flattened windowed
power spectrum rather than the deficit of power that would occur if the
quasiperiodicity was spurious. The sharpness of the low frequency peak
also argues against this being the result of a step in the power spectrum
coupled with low frequency power removal caused by detrending although we
did not investigate this possibility in depth. 

The coherence of the low frequency quasiperiodic oscillation was investigated
by fitting a sinsuoid + quartic polynomial to the timing residuals using a 
linear least squares technique. The removal of this sinusoid also removed most 
of the excess power in the peak in the flattened windowed power spectrum (as 
well as in the Deeter polynomial power spectrum) indicating that the process 
is more nearly periodic than quasiperiodic. Another indication of the near 
periodicity is the comparison of the fitted sine to the 1st derivative of the 
Crab pulse phase timing residuals which revealed four clear oscillations after 
TJD 5500 (see figure 7). Before TJD 5400 a possible phase shift and 
amplitude change may have occured showing that the oscillation process 
may be incoherent on time scales longer than 2000 days. Inspection of the 
1st derivative plot also revealed a possible small glitch occuring near 
TJD 5243 based on the occurence of a large amplitude spike that resembles 
the large amplitude spike caused by the known small glitch at TJD 6664. 
However, the data sampling around this second possible small glitch is 
too sparse to make this conclusive (see figure 7). The occurrence of the
glitch(s) does not seem to be related to the phase change in the long
term oscillation since the phase was not clearly affected by the
glitch at TJD 6664. 

The possible origin of the periodic or nearly periodic process has been 
discussed before by Lyne, Pritchard \& Smith (1988). 
Explanations include oscillations of the superfluid vortex lattice
(Tkachenko oscillations), free precession of the neutron star or the
presence of an orbiting companion. The 568 day period is roughly comparable 
to the 1000 day timescales observed in a few radio pulsars that may be 
exhibiting free precession (e.g Stairs, Lyne \& Shermar, 2000). If the 
neutron star is freely precessing then long period pulse shape changes should 
be observable as well but no significant detections of this effect have been 
reported to date.
If an orbiting companion is causing the observed rotation oscillations
we can derive limits on orbital parameters. Assuming a binary companion 
with a circular orbit, we can determine an 
$a_x \sin i = (8.65 \pm 1.0) \times 10^{-3} $ $\rm lt-sec$ from 
the cosine amplitude given in table 2 and a mass function 
\begin{equation}
f(M) \equiv \frac{4 \pi (a_x \sin i)^3}{G P_{orb}^2} 
= \frac{(M_c \sin i)^3}{(M_x + M_c)^2} =
(6.8 \pm 2.4) \times 10^{-16}\ \Msun.
\end{equation} If we assume a Crab pulsar 
mass of $M_x = 1.4\ \Msun$ and a value of $\sin i = 1$ 
we can determine a lower limit on the companion mass of 
$M_c \geq 3.2 \Mearth$ and a projected orbital radius 
$a_c \sin i = 750 \pm 93$ lt-sec or $1.50$ astronomical units. 
The presence of a planet around the Crab pulsar is an exciting possibility
but requires confirmation by detection of the coherent oscillation on longer 
timescales before this could be considered the definitive explanation.
Caution must be emphasized here as planets have often been claimed to
explain the polynomial-like timing residuals of pulsars but have later failed 
to be confirmed (e.g. Konacki et al. 1999), although it must also be 
emphasized that the longterm periodicity or quasiperiodicity here detected 
in the Crab pulsar is a real timing component and not merely an artifact of 
fitting red noise.  

Lastly, we speculate on a possible external origin for most of the
power in the $1/f^3$ timing noise component of the pulse phase. The mechanism 
causing the timing noise would have to have an amplitude that 
increases with analysis frequency, a characteristic shared by the gravitational 
perturbations caused by a body in a Keplerian orbit on its companion. Given 
that a planet may be in orbit around the Crab pulsar could there also be a disk 
of smaller bodies or planetesimals as well? The collective gravitational
perturbations from a ``disk'' of planetesimals acting on the neutron star 
will be a source of timing noise. A planetesimal of mass $m$ in a circular 
orbit with radius $r$ around a neutron star with a mass $M_x = 1.4 M_{\sun}$ 
will have a Keplerian orbital frequency (in $\rm cycles/day$) given by: 
\begin{equation}
\nu_{orb} = 36.17 r^{-\frac{3}{2}}
\end{equation} 
where $r$ is given in $\rm lt-sec$. The amplitude of the sinusoidal timing 
signature (in seconds) will be given by $\tau = (m/M_x)(r/c)$ where 
is the speed of light. Now if we assume that the average 
planetesimal mass scales with orbital radius, as might be the case if the 
planetesimals condensed out of a uniform density disk, then $m=k r$. 
Substituting into the relationship for $\tau$ and solving for $r$ in terms 
of $\nu_{orb}$ we get:
\begin{equation}
\tau = 119.7(k/M_x)(\nu_{orb})^{-\frac{4}{3}}.
\end{equation} The superposition of
timing signatures from many planetesimals over a wide range of orbits will
result in a power spectrum of the pulse phase with a slope of $-8/3$ - very
close to the observed slope of $-3$. 

We tested this idea using a Monte Carlo simulation by simulating a set of 128 
planetesimals with random initial orbital phases and a range of
orbital frequencies from $0.001 - 1\ \rm cycles/day$ distributed uniformly in
the logarithm. We fixed the constant k by choosing a planetesimal mass
at $r=750\ \rm lt-sec$. For mass distributions with $m=kr$, the power spectral 
slopes tended to be near $-3.7$, too small to fit the Crab pulsar. For a 
planetesimal mass distribution with $m=kr^{\frac{1}{2}}$ and a planetesimal 
mass of $1.62 M_{\earth}$ at $r=750\ \rm lt-sec$, the average slope of the 
power spectrum was near $-3$ over the frequency range 
$0.003 - 0.1\ \rm cycles/day$, with an average noise level in the flattened 
power spectrum of $1 \times 10^{-6}\ \rm cycles^2/day^2$ - matching that 
observed for the Crab pulsar. To fully model the Crab timing residuals the 
planetesimal at $r=750\ \rm lt-sec$ must be replaced with a planet of mass
$M \geq 3.2 \Mearth$.  
Whether such a planetesimal mass distribution and set of orbits is reasonable 
and physical is a question beyond the scope of this paper. 

The power spectrum of the timing caused by a planetesimal disk can be calculated 
analytically as well. The phase signature of the the disk of planetesimals is 
\begin{equation}
\Delta\phi(t) = \sum_k x_k \cos(2\phi\nu_k t+\psi_k)
\end{equation}
where the sum is over plantesimals $k$, and $x_k$, $\nu_k$, and $\psi_k$ are the
phase amplitude, orbital frequency, and orbital phase respectively of plantesimal
$k$. The  phase amplitude $x_k$ is given by
\begin{equation} 
x_k = \nu_o r_k \sin i\,m_k M^{-1} c^{-1}
\end{equation}
with $r_k$ the orbital radius and $m_k$ the mass of plantesimal $k$, $i$ is the
inclination of the disk, and $M$
the total system mass. The orbital radius is related to the orbital frequency by
\begin{equation}
r(\nu) = (GM)^{1/3}(2\pi\nu)^{-\twothirds} ~~. \label{rvsnu} 
\end{equation}
If we assume the orbital phases $\psi_k$ are independent then 
the autocorrelation function of $\Delta\phi$ is
\begin{equation}
A_{\Delta\phi}(\tau) = \onehalf<\sum_k x^2_k \cos(2\pi\nu_k\tau)>.
\end{equation}
If the density of plantesimals per unit radius $r$ is given by $D(r)$ and the
plantesimal mass distribution at radius $r$ is given by $P(m|r)$ then the
autocorrelation function is expected to be
\begin{eqnarray}
A_{\Delta\phi}(\tau) &=& \onehalf  
 \int D(r) P(m|r) (\nu_o r \sin i\,m M^{-1} c^{-1})^2 cos(2\pi\nu(r))dm dr 
 \nonumber \\
&=& \int \mathcal{S}(\nu)\cos(2\pi\nu\tau) d\nu 
\end{eqnarray}
where we have changed integration variables using equation \ref{rvsnu}, and
\begin{equation}
\mathcal{S}(\nu) = \onethird \pi \nu^2_o \sin^2 i G M^{-1} c^{-2} (2\pi\nu)^{-3}
D(r(\nu))m^2(r(\nu))
\end{equation}
with
\begin{equation}
m^2(r) = \int P(m|r) dm
\end{equation}
being the mean square plantesimal mass are radius $r$. 

$\mathcal{S}(\nu)$ is the
power spectrum of the phase component $\Delta\phi$. If $D(r)m^2(r)$ is a
powerlaw in r with index $\alpha$, then $\mathcal{S}(\nu)$ will be a powerlaw in
$\nu$ with index $-(3+2\alpha/3)$. The measured $\nu^{-3}$ power spectra 
implies that $\alpha \approx 0$. From the strength of the power spectrum we find
that
\begin{equation}
D(r) m^2(r) \approx 2.1 \sin^{-2} i M^2_\earth {\rm Au}^{-1}
\end{equation}
in the radius range $0.13 < r < 1.3$Au.

Remnant disks of heavy elements formed by fallback material from the supernova 
explosion that gave birth to the neutron star have been proposed as 
explanations for phenonmena observed in radio pulsars and anomalous X-ray 
pulsars (e.g. Lin et al. 1991, Alpar 2001). A potential planet or 
planets around the Crab pulsar would probably have to come from preexisting 
planets that survived the supernova rather than being assembled 
from planetesimals given the short time ($<1000$ years) since the Crab 
supernova occurred although the effect of a pulsar wind on a dust disk might 
significantly speed up the formation of planetesimals by concentrating 
the orbiting dust into preplanetesimal pockets.   
In addition to the cases of a pure fallback disk or surviving planets one can 
also envision mixed scenarios in which supernova surviving planets interact 
with a supernova fallback disk or a planetesimal disk forms through the
collision of a planet with another body.  

The presence of a planetesimal disk causing the timing noise in the Crab
pulsar and possibly in other radio pulsars with significant timing noise
would have some wider implications if validated.
The fall back scenarios proposed for anomalous X-ray pulsars (AXP's, for a
review see Mereghetti et al. 2002) have tended to consider only gaseous 
accretion disks. Large gaseous disks could supply the spin-down torque 
observed in these systems and supply accreting material to produce the 
persistant X-ray emission, but cannot explain the recently observed 
X-ray bursts (Gavriil et al. 2002) that support the idea that the related 
X-ray burst emitting soft gamma repeaters and anomalous X-ray pulsars form a 
single class of objects. In addition, searches for optical counterparts
do not support the presence of large gaseous disks in such systems (e.g.
Hulleman et al. 2000). Thus, by default, the alternative magnetar theory 
has gained much recent support.  
In contrast to a large gaseous disk, a toroidal swarm of planetesimals
would be more easily hidden from view and could supply material in the form of 
gas for steady accretion as planetesimals are tidally broken up near the 
neutron star and form an inner gaseous ring or as dust formed further out,
perhaps through plantesimal intradisk collisions, migrates inwards under
the Poynting-Robertson effect or from drag in a pulsar wind.  
Planetesimals beyond the tidal radius could also supply material for series of 
occasional X-ray bursts if a planetesimal gets fragmented as a result of an 
intradisk collision and the low angular momentum pieces gradually rain down 
onto the neutron star in an event analogous to a meteor storm. Alternatively,
a planetesimal could have its orbit disrupted into a comet-like path by a 
close encounter with one of the largest disk members and eventually collide
with the neutron star causing a giant outburst.  Series of smaller outbursts
could also occur before, during and after the main body collides with the
neutron star, if the plantesimal is broken up by a close orbital pass by 
the neutron star prior to or during the orbital pass of the primary collision. 
Either scenario would cause occasional bursts of X-rays to be emitted by the 
neutron star. Thus a plantesimal disk has the potential of explaining both 
the persistant X-ray emission observed, the slowing down of the neutron star 
spin, sets of occasional small and giant X-ray bursts and the observed lack of 
a large gaseous accretion disk. In addition, the observations of transient 
infrared emission associated with several AXP's after X-ray bursts have been 
detected (e.g. Israel et al. 2002) could be related to the formation of these 
temporary orbiting dust or debris clouds.

Chandra X-ray observations reveal an equatorial torus of material around the 
Crab pulsar (Weisskopf et al. 2000) that might be supplied by ablation and/or 
tidal disruption of plantesimals from an inner disk. The radius of the observed torus is on the order of 0.4 parsecs, 
while the disk considered here is within 1.5 astronomical units of the 
pulsar - far too small to be imaged directly. 

The fast rotation of the Crab pulsar would place it in a propellor phase
in which the outward flow of momentum from the pulsar exceeds the 
gravitational energy of the incoming material. This would
result in a particle outflow and possibly cause the observed jet from the Crab 
pulsar. The rate of spindown torque on the neutron star would also be increased
relative to the vacuum case, lowering the braking index from the 
canonical value of 3 for the emission of pure magnetic dipole radiation. 
However, to observe gravitational effects on the neutron star it is not 
necessary that the planetesimals or any derivative dust must be interacting 
directly with the pulsar magnetosphere at the present time.  

A group of planetesimals with a total mass of a few dozen Earth masses could 
plausibly cause the timing noise in the Crab pulsar and would have a mass 
similar to that required of the postulated ``fallback'' disks of the anomalous 
X-ray pulsars and soft gamma repeaters. If the scenario outlined here is 
correct, the Crab pulsar could eventually evolve into an anomalous X-ray pulsar 
or soft gamma repeater when the neutron star rotation slows to the point that 
accretion becomes possible. If other pulsars have planetesimal disks, then the 
power spectral indices of their timing noise might be expected to have a fairly
wide range, caused by differing disk mass distributions and one might observe 
real quasiperiodicities as well in their timing noise due to the largest 
bodies orbiting the neutron star.  

\acknowledgments
We thank Andrew Lyne for kindly providing the Jodrell Bank Crab timing
data and Robert Pritchard for help in the initial analysis. We also
thank Paul Boynton for useful discussions and for help in the early
stages of this project. 

\appendix

\section{Generation of Red Power-Law Noise Using Power-Law Weighting of White Noise}  

Power-law noise (PLN) processes with negative even integer spectral indices 
can be produced by the repeated integration of a white noise 
process $\epsilon(t)$. 
For example, a random walk, with a $1/f^2$ power spectrum, is defined as: 
\begin{equation} 
r_{2}(t)=\int_{-\infty}^{t} \epsilon(t^{'}) dt^{'}.
\end{equation}
We consider $\epsilon(t)$ to be a zero-mean white noise process defined for
all t with an autocovariance function
$< \epsilon(t)\epsilon(t+\tau) >=\sigma_{\epsilon}^{2}\delta(\tau)$,
where $\delta(t)$ is the Dirac delta-function. 
An equivalent way of defining $r_{2}(t)$ is as the convolution of white noise
with the unit step function, $H(t)$:
\begin{equation} 
r_{2}(t)=\epsilon(t)*H(t)=
\int_{-\infty}^{+\infty}\epsilon(t^{'})H(t-t^{'}) dt^{'}=
\int_{-\infty}^{t} \epsilon(t^{'}) dt^{'} 
\end{equation}
where $H(t)$ is the standard Heaviside step function defined as: 
$H(t)=1; \ \ t\geq 0$ and $H(t)=0; \ \ t<0 $~.
Now a $1/f^4$ noise process, produced by the integration of $r_{2}(t)$, 
can be made instead by 
convolving $H(t)$ with $r_{2}(t)$, but this is equivalent to first convolving 
$H(t)$ with itself and then convolving the resulting function with 
$\epsilon(t)$:
\begin{equation} 
H(t)*(H(t)*\epsilon(t))=(H(t)*H(t))*\epsilon(t).
\end{equation} The self convolution
of $H(t)$ results in a ramp function: $tH(t)$. Therefore $r_{4}(t)$ can
be generated directly as the convolution of $tH(t)$ with white noise:
\begin{equation} 
r_{4}(t)=(tH(t))*\epsilon(t)=\int_{-\infty}^{t}(t-t^{'})\epsilon(t^{'})
dt^{'}.
\end{equation} The convolution of 
$\epsilon(t)$ with $t^2H(t)$ will generate $1/f^6$ noise.
In general, even-integer red PLN can be generated directly
from white noise as:
\begin{equation} 
r_{m}(t)=\frac{1}{k!}\int_{-\infty}^{t} (t-t^{'})^k\epsilon(t^{'}) dt^{'} 
\end{equation}   
where the power-law spectral index $m = 2(k+1)$ 
and $m = 2,4,6 \ldots$ without the need for repeated integration. 
Note that a normalization factor of $\frac{1}{k!}$ has been added. Given this
formulation for $r_{m}(t)$, there is no {\it a priori} reason to restrict $k$ 
to positive integer values and we consider the generalization of this
formulation to noninteger $k$. 

The noise process described in equation (A5) is a member of the class of
general linear noise processes (see e.g. Priestley, 1981, p. 177) with the 
form:
\begin{equation} 
r(t)=\int_{-\infty}^{t} w(t-t^{'})\epsilon(t^{'})dt^{'} 
\end{equation} 
in which a weight function of the form $w(t)H(t)$ is convolved with white 
noise. This form has many easily analyzable properties. 
For example, the ensemble averaged PDS can be calculated 
from the multiplication of the Fourier transform of $w(t)H(t)$ with that of 
$\epsilon(t)$. Thus $r(t)$ can be described in terms of the ``noise strength''
$S$ and the weight function of the form $w(t)H(t)$ in the time domain or the 
spectral shape of the PDS in the frequency domain. In the case of discrete
white noise, the weight function can be viewed simply as a ``step'' so that 
the noise process can be viewed as an accumulation of random steps with a 
similar form. For example, a random walk is an accumulation of the step
functions H(t), occuring with random amplitudes and at random times. 
If the duration of the step is finite the noise process will be asymptotically 
stationary since two values of the noise process separated by a sufficiently 
large time interval will be uncorrelated. In the well known case of 
``shot noise'' the step is a decaying exponential, so values of the noise 
process are correlated on short timescales and approach an uncorrelated state 
on long timescales. This is reflected in the PDS of shot noise, which is flat 
at low frequencies but turns over to a power-law at high frequencies. For red
power-law noise proceses a correlation exists between a given value and all 
past values of the time-series, but the statistical properties are the same
at any time $t$ so the process is stationary.   

Let us look at a specific power-law weight function before moving onto the 
more general case. If $k=-{\frac{1}{2}}$ so 
$w(t)H(t)=\frac{1}{\sqrt{\pi}}t^{-1/2}H(t)$, where the generalized definition 
of $k! = \Gamma (1+k)$ has been used, then the Fourier transform will be 
given by: 
\begin{equation}
F(f)=\frac{1}{\sqrt{\pi}}\int_{0}^{+\infty}t^{-1/2}(\cos(2\pi ft)-i\sin(2\pi ft)) dt.
\end{equation}                                                   
The resulting Fourier transform is:
\begin{equation} 
F(f)=(1-i)(\frac{1}{2\sqrt{\pi f}}) 
\end{equation} 
with a power density spectrum of the form:
\begin{equation} 
P(f)=F(f)F^*(f)=\frac{1}{2\pi f}.
\end{equation} 
Multiplication by the PDS of $\epsilon(t)$ creates the well 
known $1/f$ noise power spectrum. 

The general power-law problem involves solving Fourier transforms of the 
weight function $W(t)=\frac{t^k}{k!}H(t)$:
\begin{equation} 
F(f)=\frac{1}{k!}\int_{0}^{+\infty}t^{k}(\cos(2\pi ft)-i\sin(2\pi ft)) dt.
\end{equation}


The Fourier integral will not in general converge since the time-series has
infinite ``energy'' but we can solve directly for a few special cases. 
For example, if $-1 < k < 0$ the integrals have the solutions:
\begin{equation} 
F(f)=
\frac{\pi}{2\Gamma(\vert k\vert)\Gamma(1+k)}\left (\frac{1}{\sin(\frac{\vert k\vert \pi}{2})}+
\frac{i}{\cos(\frac{\vert k\vert \pi}{2})} \right )
(2\pi f)^{-(k+1)} 
\end{equation} 
and the power spectrum is given by:
\begin{equation} 
P(f)=\frac{\pi^2}{4(\Gamma(\vert k\vert) \Gamma(1+k) )^2}
\left (\frac{1}{\sin(\frac{\vert k\vert \pi}{2})^2\cos(\frac{\vert k\vert \pi}
{2})^2} \right )(2\pi f)^{-2(k+1)}. 
\end{equation}
When $k=-\frac{1}{2}$ then (A11) and (A12) reduce to the previous results 
(A8) and (A9) for $1/f$ noise. 

Although the Fourier transform can't be solved directly it can be solved
as a limit. If we add an exponential cutoff we get an integral of the form:
\begin{equation}
F(f)=\frac{1}{k!}\int_{0}^{+\infty}e^{-\frac{t}{\lambda}}t^k(\cos(2\pi ft)-i\sin(2\pi ft)) dt
\end{equation} 
The integral solutions to the cosine and sine portions of the integral can be 
found in integral tables and are, respectively:
\begin{equation}\frac{1}{k!} 
\int_{0}^{+\infty}e^{-\frac{t}{\lambda}}t^k \cos(2\pi ft) dt =
\frac{\left [ (\frac{1}{\lambda}-i2\pi f)^{k+1}+
(\frac{1}{\lambda}+i2\pi f)^{k+1} \right ]}
{2(\frac{1}{\lambda^2}+(2\pi f)^2)^{k+1}} 
\end{equation}
and 
\begin{equation}\frac{1}{k!} 
\int_{0}^{+\infty}e^{-\frac{t}{\lambda}}t^k \sin(2\pi ft) dt =
\frac{\left [(\frac{1}{\lambda}+i2\pi f)^{k+1}-
(\frac{1}{\lambda}-i2\pi f)^{k+1} \right ]}
{2i(\frac{1}{\lambda^2}+(2\pi f)^2)^{k+1}}.
\end{equation}
In the limit as $\lambda \rightarrow \infty$ the Fourier transform will
approach the value:
\begin{equation} 
F(f)=(-i)^{k+1}(2\pi f)^{-(k+1)} 
\end{equation}
and the power density spectrum will approach the power-law form:
\begin{equation} 
P(f)=(2\pi f)^{-2(k+1)}. 
\end{equation}
The full ensemble averaged PDS will be given by:
\begin{equation}
P(f)=\frac{\sigma_{\epsilon}^2}{(2\pi f)^{m}}
\end{equation}
where $\sigma_{\epsilon}^2$ is the mean power level of the underlying white 
noise (equal to the noise strength S) and $m=2(k+1)$ is the power-law 
spectral index. Now equation A5 can be viewed as a general formula for
producing power-law noise by the convolution of a step $t^kH(t)$ with
a white noise process.

\subsection{Time Domain Properties} 

In practice, any physically realizable red noise time series will first be 
observed at a time $t=0$ and have started at this or some earlier time. The 
linearity of the convolution integral allows the time series to be divided 
into the form:
\begin{equation} 
r_{m}(t)=\frac{1}{k!}\int_{T}^{0} (t-t^{'})^k\epsilon(t^{'}) dt^{'} + 
         \frac{1}{k!}\int_{0}^{t} (t-t^{'})^k\epsilon(t^{'}) dt^{'}
\end{equation}
where $T$ can range to $-\infty$ and $m=2(k+1)$.
The first integral referring to the past will contribute a smooth function 
to the time series after $t=0$. Note that in this form, $r_{m}(t)$ is
divided into two nonstationary processes. When $\epsilon(t)$ is viewed
as a series of discrete impulses the time series $r_{m}(t)$ can be viewed
as an accumulation of ``steps'' of the form 
$s(t) = t^kH(t)$. The smooth function will have a form related to that
of an individual step. 
If the index $k$ is a positive integer, the 
smooth function will have the form of a simple polynomial. For the random 
walk, k=0 ($1/f^2$ noise), the time series consists of a series of randomly
ascending or descending ``stair steps'' of variable amplitude and duration. 
The smooth function will be a constant: 
\begin{equation} 
\int_{T}^{0} \epsilon(t^{'}) dt^{'}=C_0.    
\end{equation}
For k=1 ($1/f^4$ noise), the time series consists of a series of connected
linear segments (``ramps'') and the smooth function will be a linear trend:
\begin{equation}
\int_{T}^{0} (t-t^{'})\epsilon(t^{'}) dt^{'}=t\int_{T}^{0} \epsilon(t^{'})
dt^{'} - \int_{T}^{0} t^{'}\epsilon(t^{'}) dt^{'}=C_0t-C_1
\end{equation}  
and for $k=2$ ($1/f^6$ noise), the time series consists of quadratically shaped
segments and the smooth function will be a quadratic trend and so on. Note 
that a polynomial fit of an order $l \ge k$ will remove the smooth function
component completely when $k$ is an integer. 

The ensemble averaged expectation and variance of power-law noise at a time t 
can be easily calculated. Each point of a power-law noise realization
simply consists of a linear combination of white noise, so the statistics 
will be directly related to the white noise statistics.
The expectation, $<r_m(t)>$ will be given by:
\begin{equation}
<r_m(t)>=\frac{1}{k!}\left <\int_{-\infty}^t (t-t^{'})^k\epsilon(t^{'}) dt^{'} \right > =
\left <\epsilon(t) \right >\frac{1}{k!}\int_{-\infty}^t (t-t^{'})^k dt^{'}.
\end{equation}
If the times series is divided into two finite portions as in A19 
and $k \neq -1$ the expectation has the simple form:
\begin{equation}
<r_m(t)>=\frac{<\epsilon(t)>}{(k+1)!}(t+T)^{k+1}.
\end{equation}
A zero mean white noise process implies that $<r_m(t)>=0$ as well. 

The variance of $r_m(t)$ as a function of time, 
$<r_m(t)^2>-<r_m(t)>^2$ can be calculated from:
\begin{eqnarray}
<r_m(t)^2> & = & \left (\frac{1}{k!}\right )^2
\left <\left (\int_{{-\infty}}^{t} (t-t^{'})^k\epsilon(t^{'}) dt^{'}\right )^2 
\right > \nonumber \\
 & = & \left (\frac{1}{k!}\right )^2 \int_{{-\infty}}^{t}
\int_{{-\infty}}^{t} (t-t^{'})^k (t-t^{''})^k \left < \epsilon(t^{'})
\epsilon(t^{''}) \right > dt^{'}dt^{''}
 \nonumber \\
 & = & \left (\frac{1}{k!}\right )^2 \int_{{-\infty}}^{t} (t-t^{'})^{2k}\left 
< \epsilon(t^{'})^2 \right > dt^{'} \nonumber \\
 & = & \left (\frac{\sigma_{\epsilon}}{k!}\right )^{2} \int_{{-\infty}}^{t} (t-t^{'})^{2k} dt^{'}
\end{eqnarray}
where use is made of the fact that for white noise:
\begin{equation}
\left <\epsilon(t^{'})\epsilon(t^{''})\right >=
\sigma^2_{\epsilon} \delta(t^{'}-t^{''})
\end{equation} where $\delta(t)$ is the Dirac delta-function.               
Note that the variance $<r_m(t)^2>$ at time t is in fact infinite for all
values of k.
If the time series is divided into two finite portions as in A19 
with zero mean white noise being used and $k \neq -0.5$, the variance is 
finite and reduces to the simple expression:
\begin{equation}
<r_m(t)^2>=\frac{1}{(2k+1)}
\left (\frac{\sigma_{\epsilon}}{k!}\right )^{2}(t+T)^{2k+1}.
\end{equation}
Therefore, if one could measure $<r_m(t)^2>$ in this way, by averaging
over a number of noise realizations for example, the
PDS spectral index $m = 2(k+1)$, and the underlying white noise variance 
$\sigma_{\epsilon}^{2}$ could be estimated in a straight forward manner.

The autocovariance can also be derived in a similar manner. The autocovariance 
is, for $<r_m(t)>=0$, given by:
\begin{equation}
<r_m(t)r_m(t+\tau)>=\left (\frac{1}{k!}\right )^2
\left <\left (\int_{{-\infty}}^{t} (t-t^{'})^k\epsilon(t^{'})dt^{'}\right ) 
\left (\int_{{-\infty}}^{t+\tau} (t+\tau-t^{'})^k\epsilon(t^{'})dt^{'}\right ) 
\right >.                                                                      
\end{equation}
This expression reduces to:
\begin{equation}
<r_m(t)r_m(t+\tau)>=\left (\frac{\sigma_{\epsilon}}{k!}\right )^{2}
\left (\int_{{-\infty}}^{t} (t-t^{'})^k(t+\tau-t^{'})^k dt^{'}\right )
\end{equation}
with $\tau \geq 0$.
Note that the autocovariance function depends on both the current time $t$ as
well as the lag $\tau$. The Fourier transform of the autocovariance function
has the form:
\begin{equation}
\int_{-\infty}^{\infty} 
e^{-i2\pi f\tau} <r_m(t)r_m(t+\tau)> d\tau 
\end{equation}
The exponential can be broken into the two terms 
$e^{-i2\pi f\tau} = e^{-i2\pi f(\tau -(t^{'}-t))} e^{-i2\pi f(t^{'}-t)}$ and 
after rearranging, the integral divides into two terms, one of the form A10 
and one of its conjugate that can be solved in the limit as in A13. The 
Fourier transform will, in the limit, approach the power-law form given 
by A18.

Expression A27, after removing the past smooth function, has the form:
\begin{equation}
<r_m(t)r_m(t+\tau)>=\left (\frac{\sigma_{\epsilon}}{k!}\right )^{2}
\left (\int_{0}^{t} (t-t^{'})^k(t+\tau-t^{'})^k dt^{'}\right )
\end{equation}
which reduces to A24 when $\tau=0$.
For the three cases in which k=0, 1 and 2, i.e. for $1/f^2$, $1/f^4$ and $1/f^6$ 
PLN this expression has the following forms: \\
For k=0:
\begin{equation}
<r_{2}(t)r_{2}(t+\tau)>=\sigma_{\epsilon}^{2}t
\end{equation}
for k=1:
\begin{equation}
<r_{4}(t)r_{4}(t+\tau)>=\frac{1}{6} \sigma_{\epsilon}^{2}t^2 
\left ( 2t + 3\tau \right )                              
\end{equation}  
and for k=2:
\begin{equation}
<r_{6}(t)r_{6}(t+\tau)>=\frac{1}{120} \sigma_{\epsilon}^{2}t^3
\left ( 6t^2+15t\tau + 10\tau^2 \right )
\end{equation}
Note the time dependence of the autocovariance for $r_{4}(t)$ and $r_{6}(t)$
after removal of the past smooth function. When the substitutions
$t=t_{<}$ and $t+\tau=t_{>}$ are used then A31, A32 and A33 are seen to be 
the continuous white noise version of equations (23), (38) and (43) derived by 
Groth (1975b).

%

\subsection{Discrete Red Power-law Noise}

Any realizable red PLN process, in addition to being of limited 
duration, will be composed of steps of a finite size occurring at a finite 
rate. We now consider a discrete model of red PLN that is useful for 
generation of actual samples with Monte-Carlo routines.
A discrete form of (A5) can be obtained by replacing $\epsilon (t)$ with
the discrete white noise form:
\begin{equation} 
\epsilon (t) = \sum_{j=-\infty}^{\infty}\epsilon_j\delta(t_j)
\end{equation} which produces the discrete power-law noise process:
\begin{eqnarray}
\hat{r}_{m}(t) & = & 
\frac{1}{k!}\sum_{j=-\infty}^{j_{max}}(t-t_j)^k H(t-t_j) \epsilon(t_j)
\end{eqnarray}  
where the power-law spectral index $m = 2(k+1)$ and $H(t)$ is the unit step function as 
defined previously. The last white noise ``event'' in the time-series,
$\epsilon(t_j)$, occurs at a time $t_{j_{max}}$ such that 
$t_{j_{max}} \leq t < t_{{j_{max}}+1}$. We choose the duration between events
$\Delta t = t_j-t_{j-1}$ to be a random variable with a probability density 
function: $p(\Delta t)= R_g e^{-R_g \Delta t}$, i.e. the ``waiting time'' 
distribution of the Poisson process. The $\epsilon(t_j)$ are drawn from a 
discrete zero mean gaussian white noise process with a variance 
$\sigma_{\epsilon}^{2}(\Delta t_g)=\Delta t_g \sigma_{\epsilon}^2$
and a mean rate of step generation $R_g=\frac{1}{\Delta t_g}$ such that 
$\Delta t_g = <\Delta t>$ (a noise process sometimes referred to as
{\it white Poisson impulse noise}). The noise strength is given by 
$S=R_g\sigma_{\epsilon}^2(\Delta t_g)$ and along with the index k and $R_g$ 
completely specify the type of noise process. Processes with the same
noise strength can be made by the frequent accumulation of many small
steps or with fewer, larger steps in the same time interval. 

The physical units 
of $S$ depend on the units of $\hat{r}_{m}(t)$ and the specific power-law. 
For example, let $\hat{r}_{m}(t)$ be noise in the rotation phase of a pulsar 
so the physical units are cycles and let time be given in seconds. Then the 
units of $\epsilon(t_j)$ will be $\rm cycles\ sec^{-k}$. 
Because $\sigma_{\epsilon}(\Delta t_g)=\sqrt{\frac{S}{R_g}}$ the units 
of $S$ are $\rm cycles^2 \ sec^{-(1+2k)}$. If $\hat{r}_{m}(t)$ is noise in the 
rotation frequency then the units of $\epsilon(t_j)$ will be 
$\rm Hz \ sec^{-k}$ and the units of $S$ will be $\rm Hz^2 \ sec^{-(1+2k)}$. 
If noise in the rotation frequency derivative then $S$ will have units of 
$\rm (Hz/sec)^2 \ sec^{-(1+2k)}$, etc. 
 
Any physical or simulated time-series will begin at a time that we can
arbitrarily set at $t=0$. Then $\hat{r}_{m}(t)$ will have the form:
\begin{equation}
\hat{r}_{m}(t)=\frac{1}{k!}\sum_{j=0}^{j_{max}}(t-t_j)^k H(t-t_j) \epsilon(t_j).
\end{equation}  
In this form it can be seen that $\hat{r}_{m}(t)$ is simply a generalization
of the discrete PLN models contained in Groth (1975b), in 
which k=0, 1,and 2 correspond to his phase, frequency and slow-down noise
definitions respectively. We can also rewrite $\epsilon(t_j)$ as a scaled 
version of a unit variance time-series $\epsilon_{u}(t_j)$ such that: 
$\epsilon(t_j)=\sqrt{S \Delta t_g} \epsilon_{u}(t_j)$ where 
$\sigma_{\epsilon_{u}}^{2}(\Delta t_g)=1$. Equation (A35) then becomes:
\begin{equation}
\hat{r}_{m}(t)=\frac{\sqrt{S \Delta t_g}}{k!}\sum_{j=0}^{j_{max}}(t-t_j)^k H(t-t_j) \epsilon_u(t_j).
\end{equation}   
and is useful for generating discrete noise realizations of noise strength S. 
Now if $\hat{r}_{m}(t)$ above is used in its entirety to represent a power-law
noise realization, the low frequency power contributed by the past will be
too small so for an accurate representation, a sufficiently long portion of 
the early time series needs to be discarded.

Following Groth (1975b), we now derive the mean $<\hat{r}_{m}(t)>$ and 
autocovariance $<\hat{r}_{m}(t)\hat{r}_{m}(t+\tau)>$ for equation (A36).
Now $<\hat{r}_{m}(t)>$ is given by:
\begin{eqnarray}
<\hat{r}_{m}(t)> & = & \left <\frac{1}{k!}\sum_{j=0}^{j_{max}}(t-t_j)^k H(t-t_i)
\epsilon(t_j) \right > \nonumber \\
 & = & \frac{1}{k!}\sum_{j=0}^{j_{max}} <(t-t_j)^k H(t-t_j)><\epsilon(t_j)> \nonumber \\
 & = & \frac{<\epsilon(t)>}{k!}\sum_{j=0}^{j_{max}} <(t-t_j)^k H(t-t_j)> \nonumber \\
 & = & \frac{<\epsilon(t)>}{k!}R_g\int_{0}^{t} (t-t^{'})^k dt^{'} \nonumber \\
 & = & \frac{<\epsilon(\Delta t_g)>}{(k+1)!}t^{k+1} \ \ \ (k \neq -1)  
\end{eqnarray}
where $<\epsilon(\Delta t_g)>$ is the mean on a time-scale 
$\Delta t_g = R_g^{-1}$. This is the same equation as (A23) and when k=0, 1, 
and 2 reduces to Groth's results for phase, frequency and slow-down noise.

The autocovariance is given by:
\begin{eqnarray}
<\hat{r}_{m}(t)\hat{r}_{m}(t+\tau)> & = & 
\left < \left ( \frac{1}{k!}\sum_{j=0}^{j_{max}}(t-t_j)^k H(t-t_j) \epsilon(t_j) \right )
\left ( \frac{1}{k!}\sum_{j=0}^{j_{max}+l}(t+\tau-t_j)^k H(t+\tau-t_j) \epsilon(t_j) \right )
 \right
> \nonumber \\
 & = & \left (\frac{1}{k!} \right )^2 \sum_{j=0}^{j_{max}}\sum_{j^{'}=0}^{j_{max}+l}
\left < (t-t_j)^k(t+\tau-t_{j^{'}})^k H(t-t_j) H(t+\tau-t_{j^{'}}) \right > 
\left <\epsilon(t_j)\epsilon(t_{j^{'}}) \right > \nonumber \\
 & = & \left (\frac{\sigma_{\epsilon}(\Delta t_g)}{k!} \right )^2 \sum_{j=0}^{j_{max}} \left <
(t-t_j)^k(t+\tau-t_j)^k H(t-t_j) H(t+\tau-t_{j}) \right > \nonumber \\
 & = & R_g \left (\frac{\sigma_{\epsilon}(\Delta t_g)}{k!} \right )^2 
\int_{0}^{t} (t-t^{'})^k(t+\tau-t^{'})^k dt^{'} 
\end{eqnarray}  
where $t_{j_{max}} \leq t < t_{{j_{max}+1}}$ and 
$t_{j_{max}+l} \leq t+\tau < t_{j_{max}+l+1}$. The autocovariances derived by 
Groth (1975b) for his phase, frequency and slow-down noise definitions 
(k=0, 1 and 2 respectively) can be obtained from equation (A39) by the 
substitutions $t_{>}=t+\tau$ in the second term in the integral and 
$t_{<} = t$ elsewhere.

\section{Generation of Red Power-Law Noise Using a Modified Timmer-K\"{o}nig
Method}  

A method for simulating power-law noise realizations has been proposed by 
Timmer \& K\"{o}nig (1995). Real and imaginary values of the  
Fourier amplitudes corresponding to a set of discrete Fourier frequencies 
are generated from a zero mean gaussian distribution and 
then weighted by a chosen Fourier frequency dependent power-law. The 
amplitudes are then Fourier transformed into the time-domain to generate a 
uniformly sampled time-series. A normalization for a particular noise strength 
is not given in their paper. We present below, in a recipe form, a modification of this method that can be 
used to generate power-law noise realizations with noise strength $S$. 

To generate a power-law noise realization with $N$ points we created two 
zero mean unit variance gaussian white noise vectors $x_j$ and $y_j$, 
$j=0,1 \ldots N-1$ and a corresponding weighting 
vector $w_j = (\frac{1}{j})^{\frac{m}{2}}$ where $m$ is the desired 
power-law index. For the conventions used below, we assume that $N$ is an even 
integer. We then formed a complex vector of Fourier amplitudes 
$r = A\left[0,(w_jx_j,w_jy_j),(w_{N-1-j}x_{N-1-j},w_{N-j-1}y_{N-1-j})^*
\right]$ where the first member of a complex pair $(w_jx_j,w_jy_j)$ is real and 
the second imaginary and the asterisk indicates a complex conjugate. 
The third set of terms contains the negative Fourier frequencies in which 
$j=1 \ldots N-1$, while for the second set of positive Fourier frequencies, 
$j=0 \ldots N-1$, so that the complex vector $r$ contains a total of $2N$ 
terms. The Nyquist term was made purely real so that $y_{N-1}=0$. The 
normalization constant 
$A = (SN)^{\frac{1}{2}}(\frac{N}{\pi})^\frac{m}{2}$ and was chosen so that
the generated realization has an average noise strength S. A realization
was then generated as a time series via the discrete inverse Fourier
transform:
\begin{equation}
R(t_k) = \frac{1}{2N} \sum_{l=0}^{2N-1} r_l e^{\frac{-i\pi lk}{N}} \ \
k = 0,1, \ldots 2N-1.
\end{equation}
While $R(t_k)$ is a time-series of N independent points, only the first
$N/2$ should be used. The autocorrelation function of the times-series 
produced by equation B1 is approximately correct for lags less than $N/2$, 
but is wrong for larger lags. This is due to the cyclic nature of the discrete 
Fourier transform, which treats the last point as the nearest neighbor of the 
first. Use of the full time series results in an under representation of low 
frequency power. Further discussion can be found in section 7 of this paper.


\input{psfig.tex}





\newpage

\makeatletter
\def\jnl@aj{AJ}
\ifx\revtex@jnl\jnl@aj\let\tablebreak=\nl\fi
\makeatother


\begin{deluxetable}{lrrrcrrrr}
\tablewidth{33pc}
\tablecaption{Crab pulse phase ephemeris}
\tablehead{
\colhead{Parameter}           & \colhead{Value}}
\startdata          

$t_0(MJD)$                                       &  $45395.94232638889$  \nl
$\phi_0(cycles)$                                 &  $4.445(6)$           \nl
$\nu_0(Hz)$                                      & $30.04669082632(13)$ \nl
$\dot \nu_0(10^{-10}\ Hz\ s^{-1})$               & $-3.80599313(13)$    \nl
$\ddot \nu_0(10^{-20}\ Hz\ s^{-2})$              &  $1.214636(13)$      \nl
$\stackrel{\ldots}{\nu_0}(10^{-30}\ Hz\ s^{-3})$ & $-1.425(16)$  \nl


\enddata
\end{deluxetable}                                 

\newpage

\makeatletter
\def\jnl@aj{AJ}
\ifx\revtex@jnl\jnl@aj\let\tablebreak=\nl\fi
\makeatother


\begin{deluxetable}{lrrrcrrrr}
\tablewidth{33pc}
\tablecaption{Fitted Cosine parameters}
\tablehead{
\colhead{Parameter}   & \colhead{Value}}
\startdata          


$t_0(MJD)$            &  $46390.0$         \nl
$A(cycles)$           &  $0.23 \pm 0.03$   \nl
$B(cycles)$           &  $0.12 \pm 0.03$   \nl
$\Phi_0(radians)$     &  $1.09 \pm 0.116$  \nl
$P(days)$             &  $568 \pm 10$      \nl   
$A_0(cycles)$         &  $0.26 \pm 0.03$   \nl

\enddata
\end{deluxetable}                                 


\newpage

\figcaption{
Top: Crab arrival time residuals obtained from radio observations by
Jodrell Bank after removal of a quartic polynomial and correcting
for dispersion effects. 
Middle: Power spectrum of the frequency derivative computed from the
above residuals using the Deeter polynomial method (solid circles). Measurement 
noise power is shown by the triangles. 
Bottom: Deeter polynomial power spectrum (solid circles) after dividing by $2\pi f$.
The measurement noise power (triangles) has been similarly corrected. The
histogram displays the windowed power spectrum of the arrival time residuals
displayed in the top panel using a Hann window with $\alpha=2$. 
\label{1}}

\figcaption{
Examples of discrete red-noise time series showing resolved steps. The
number on the far right gives the power-law index of the process. The
bottom time series shows the discrete white noise realization from which
the red-noise time series were generated. 
\label{2}}  

\figcaption{
Sample of red noise realizations of increasing redness generated
from the same underlying white noise realization (left panel set).
The power-law index, $m$, for $1/f^m$ noise is given in the inset on 
each left panel set. The right panel set shows the same realizations
after detrending by a cubic polynomial fit.                     
\label{3}}

\figcaption{
Top: Example of a cubic polynomial detrended $1/f^3$ noise realization
with a noise strength $S = 1 \times 10^{-6}$ $\rm cycles^2/day^2$
(approximately equal to the Crab pulsar)
generated using the time-domain method. 
Middle: Hann windowed ($\alpha=2$) power density spectrum (PDS)
showing power-law slope of $-2.998 \pm 0.077$ from fit to power
between vertical dashed lines. For comparison, the smoother curve offset
above is an average PDS of 1024 identically generated 
realizations. 
Bottom: Flattened PDS after multiplication by $(2\pi f)^3$, where f is
frequency. The smoother average flattened PDS is also shown offset above
where the highest frequency power exhibits an increase by $2\times$ due 
to aliasing.                                                          
\label{4}}  

\figcaption{
Averages of 1024 flattened power spectra for power-law noise realizations
of unit strength. A high frequency sinusoid (visible as peak 
near $f=0.1$ and white noise have been added to each realization. 
Power-law spectral index $m$ shown to left of
the flat lines that display the predicted pure power-law form for unit noise 
strength. Power spectra with $m > 2.0$ have been offset in power for display 
purposes, otherwise scale is the same. The average spectra of the white noise 
alone (for $f > 0.08$) is shown as the sloping lines. 
The lower spectra of each pair show average power spectra after realizations have been 
detrended by cubic polynomials.                                          
\label{5}}

\newpage
\figcaption{
Top: Crab arrival time residuals obtained from radio observations by 
Jodrell Bank after removal of a quartic polynomial, correcting
for dispersion effects {\it and} interpolating onto a uniform grid.  
Middle: Hann windowed ($\alpha = 1$) power density 
spectrum showing power-law slope of $-3.09 \pm 0.05$ from fit to power
between vertical dashed lines (frequency range $0.003 - 0.1$ $\rm cycles/day$). 
Horizontal line shows calculated power level of statistical measurement noise. 
Bottom: flattened PDS after multiplication by $(2\pi f)^3$, where f is
frequency. Note the highly significant peak at $f = 0.0018$ $\rm cycles/day$.
Mean power level (between vertical dashed lines) is $9.26(0.50)\times 10^{-7}$
$\rm cycles^2/day^2$. Solid vertical line shows cosine frequency from table 2.
\label{6}}                                                       

\figcaption{
Top: The numerical 1st derivative of the uniformly binned spline fit to
the Crab timing residuals. The superposed sinusoid is the derivative of
the sinusoid fit to the Crab timing phase residuals. 
Middle: A closeup of the Crab timing residuals at the time of the small
glitch occuring at JD -- 2440000.5 = 6664. The raw mesurements are plotted
in top, the spline interpolation underneath. Errors are plotted for the
raw measurements but are generally smaller than the symbol size. 
Bottom: A similar closeup of the region associated with the first large 
excursion in the numerical 1st derivative. A possible glitch may
have occured here.                                                      
\label{7}}                                                      

\figcaption{
Comparison of average power spectra computed using the Deeter polynomial method
(diamonds with error bars) and windowed power spectra using a Hann window
with $\alpha=2$ of the same 32 uniformly sampled noise realizations. The noise 
is $1/f^3$ PLN generated using the time domain method and detrended by a cubic 
polynomial fit. The spectra have been flattened for the case of a $1/f^3$ 
power-law. High (top two spectra) and low (bottom three spectra) frequency 
sinusoids have been added to the noise realizations to show the effect of 
discrete features on the power spectrum estimation. The power spectra have 
been shifted in power for display purposes but the noise strength is the same 
in all cases.                                                                  
\label{8}}                          

\figcaption{
Top: Crab pulsar optical timing pulse phase residuals from Groth (1975c)
for period after 1969 glitch. Offset below is a cubic spline interpolation
onto a uniform time series.
Middle: Solid circles mark the Deeter polynomial power spectrum of the pulse 
frequency derivative (flat for $1/f^4$ noise in pulse phase). Triangles 
connected by  dotted line mark computed measurement noise spectrum. 
Dashed line shows power level measured by Deeter (1981) in his analysis
of these data. 
Bottom: Above PDS after multiplication by $(2\pi f)$, which assumes that
the real spectrum is $1/f^3$ noise in pulse phase. Note the ``bump'' in
the power near log frequency $-3.0$. Dashed horizontal line shows noise power 
level measured from Jodrell Bank measurements. Solid vertical line marks
a period of 568 days. Histogram shows Hann windowed power 
($\alpha=2$) power of spline interpolation flattened assuming an power-law 
index of $-3$.                                                  
\label{9}}                           



\newpage

\centerline{\hbox{
\psfig{figure=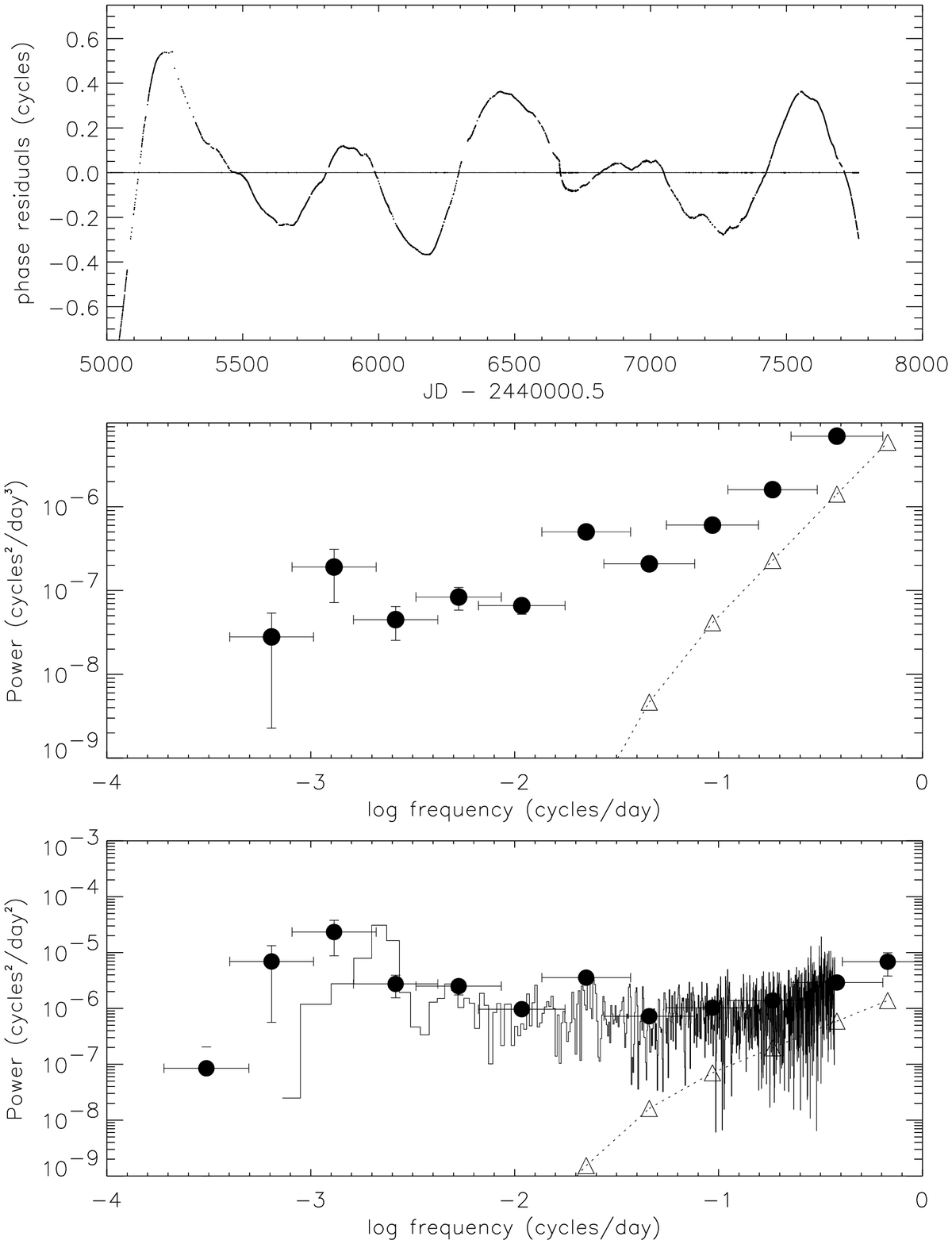,height=6.0in,width=4.6in} }}
\hfill
\parbox{5.5in}{
Fig 1. Top: Crab arrival time residuals obtained from radio observations by
Jodrell Bank after removal of a quartic polynomial and correcting
for dispersion effects. 
Middle: Power spectrum of the frequency derivative computed from the
above residuals using the Deeter polynomial method (solid circles). Measurement 
noise power is shown by the triangles. 
Bottom: Deeter polynomial power spectrum (solid circles) after dividing by $2\pi f$.
The measurement noise power (triangles) has been similarly corrected. The
histogram displays the windowed power spectrum of the arrival time residuals
displayed in the top panel using a Hann window with $\alpha=2$.
}

\newpage  

\centerline{\hbox{
\psfig{figure=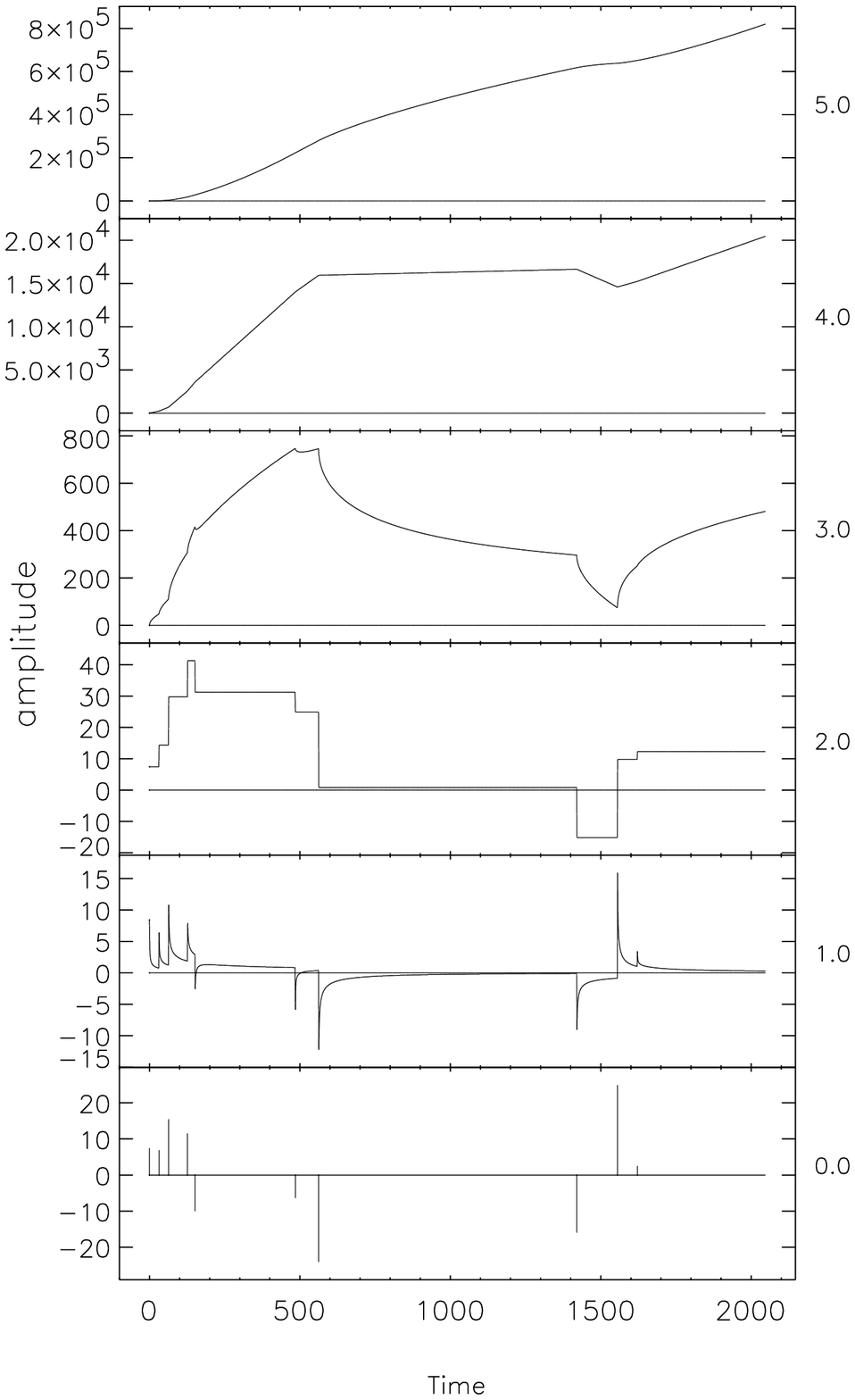,height=6.0in,width=4.6in} }}
\hfill
\parbox{5.5in}{
Fig 2. Examples of discrete red-noise time series showing resolved steps. The
number on the far right gives the power-law index of the process. The
bottom time series shows the discrete white noise realization from which
the red-noise time series were generated.
}                  

\newpage

\centerline{\hbox{
\psfig{figure=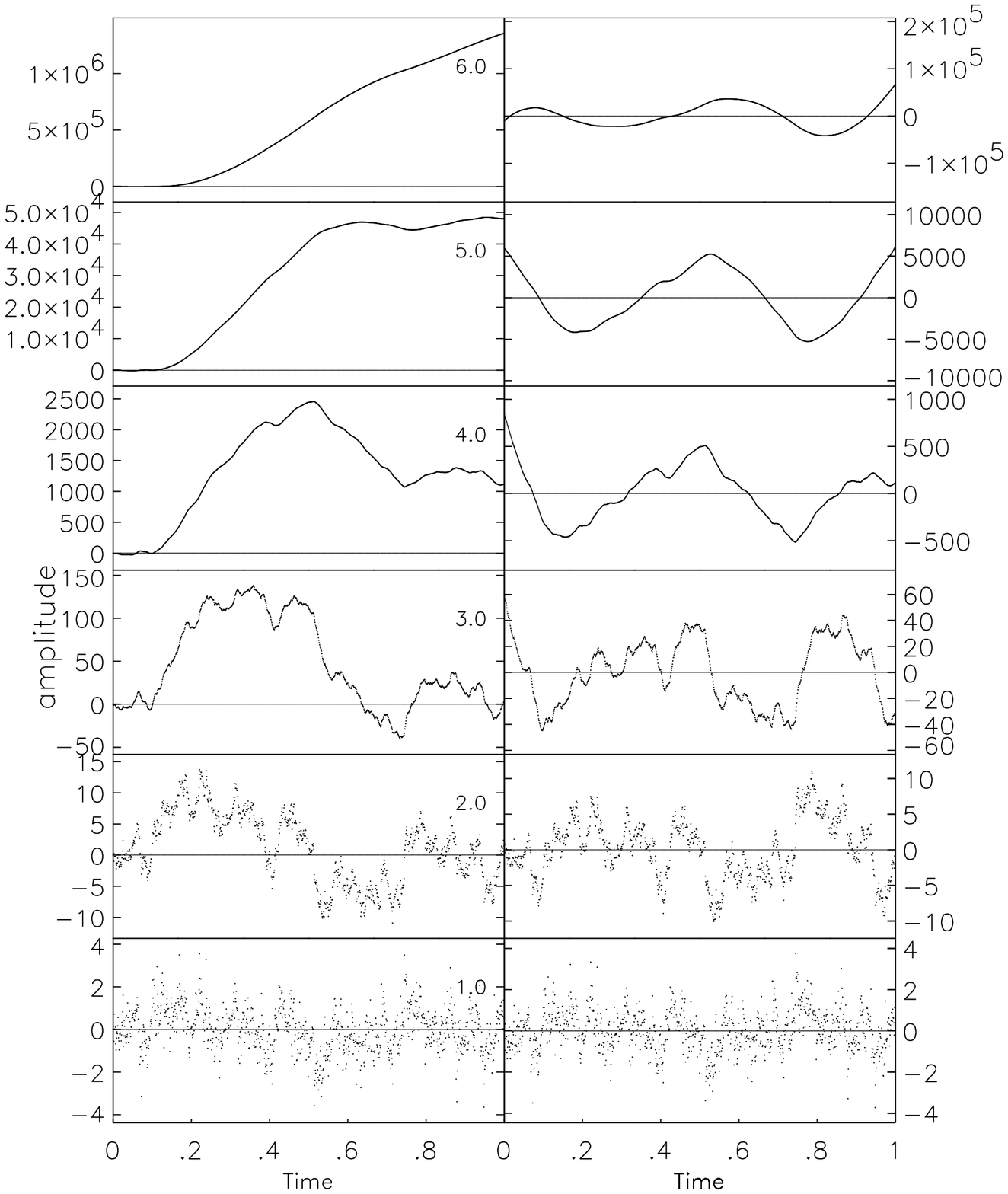,height=6.0in,width=4.6in} }}
\hfill
\parbox{5.5in}{
Fig 3. Sample of red noise realizations of increasing redness generated
from the same underlying white noise realization (left panel set).
The power-law index, $m$, for $1/f^m$ noise is given in the inset on 
each left panel set. The right panel set shows the same realizations
after detrending by a cubic polynomial fit.                      
}               

\newpage

\centerline{\hbox{
\psfig{figure=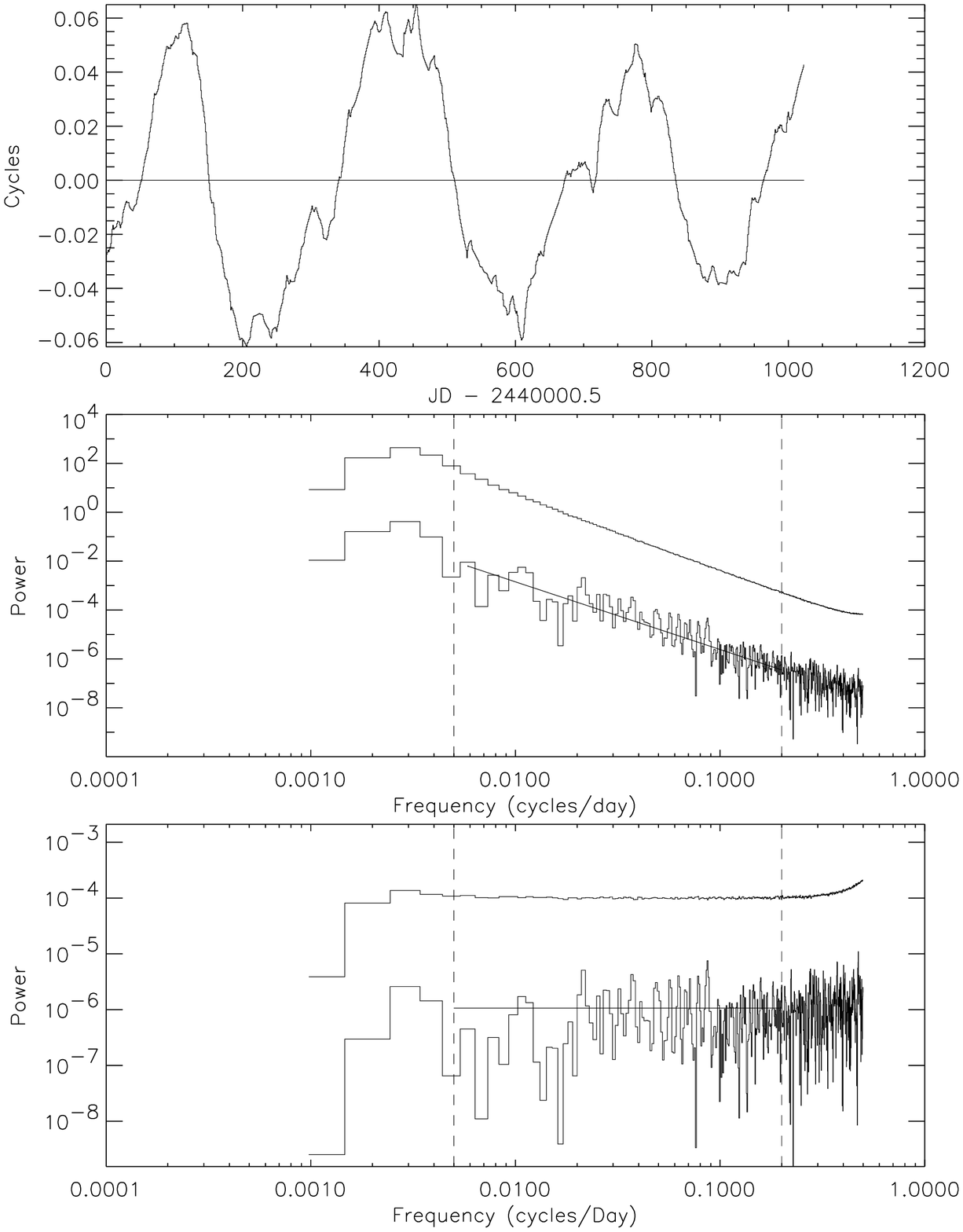,height=6.0in,width=4.6in} }}
\hfill
\parbox{5.5in}{
Fig 4. Top: Example of a cubic polynomial detrended $1/f^3$ noise realization
with a noise strength $S = 1 \times 10^{-6}$ $\rm cycles^2/day^2$
(approximately equal to the Crab pulsar)
generated using the time-domain method. 
Middle: Hann windowed ($\alpha=2$) power density spectrum (PDS)
showing power-law slope of $-2.998 \pm 0.077$ from fit to power
between vertical dashed lines. For comparison, the smoother curve offset
above is an average PDS of 1024 identically generated 
realizations. 
Bottom: Flattened PDS after multiplication by $(2\pi f)^3$, where f is
frequency. The smoother average flattened PDS is also shown offset above
where the highest frequency power exhibits an increase by $2\times$ due 
to aliasing.                                                          
}






\newpage

\centerline{\hbox{
\psfig{figure=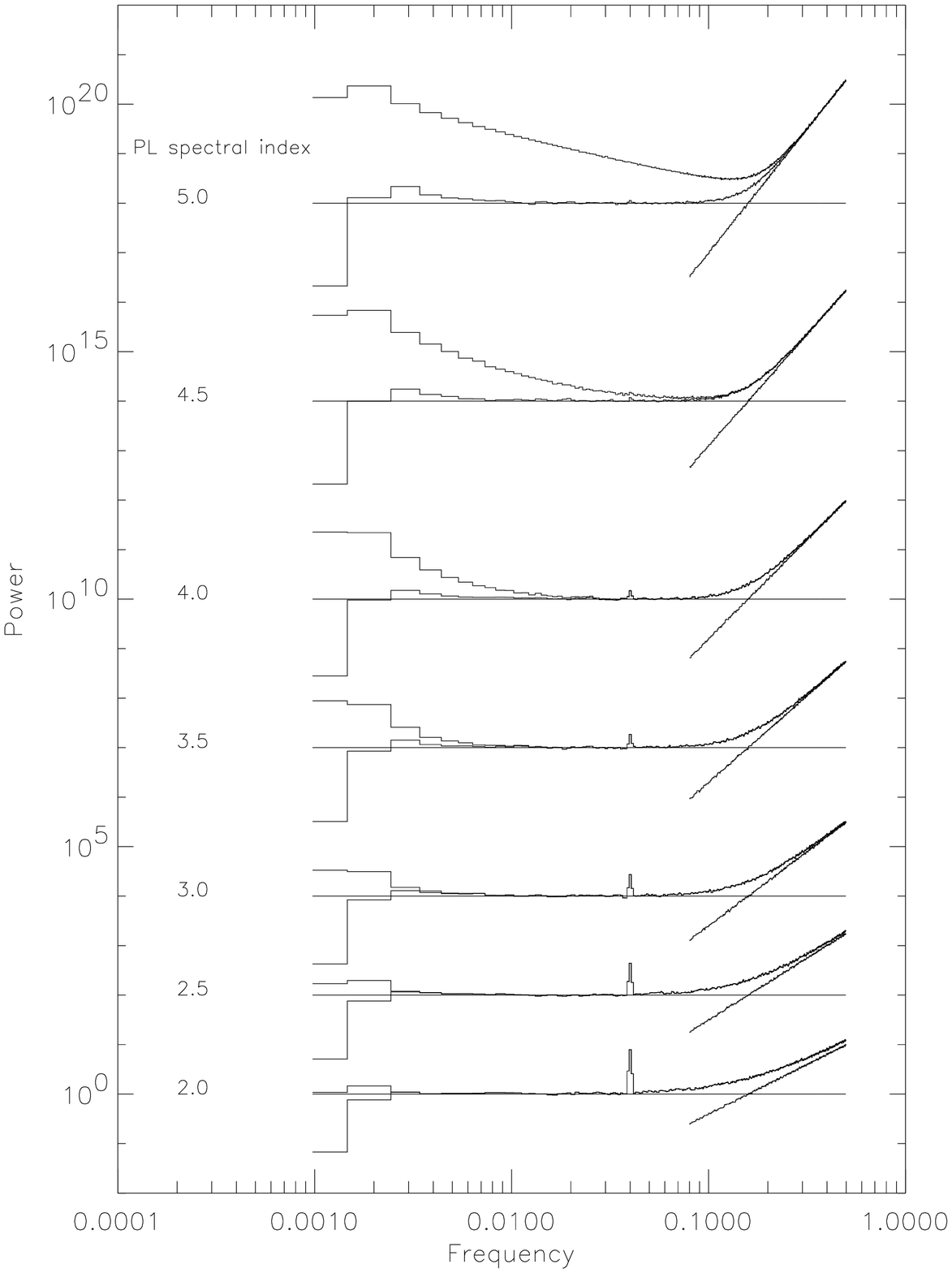,height=6.0in,width=4.6in} }}
\hfill
\parbox{5.5in}{
Fig 5. Averages of 1024 flattened power spectra for power-law noise realizations
of unit strength. A high frequency sinusoid (visible as peak 
near $f=0.1$ and white noise have been added to each realization. 
Power-law spectral index $m$ shown to left of
the flat lines that display the predicted pure power-law form for unit noise 
strength. Power spectra with $m > 2.0$ have been offset in power for display 
purposes, otherwise scale is the same. The average spectra of the white noise 
alone (for $f > 0.08$) is shown as the sloping lines.  
The lower spectra of each pair show average power spectra after realizations have been 
detrended by cubic polynomials.                                          
}  

\newpage       

\centerline{\hbox{
\psfig{figure=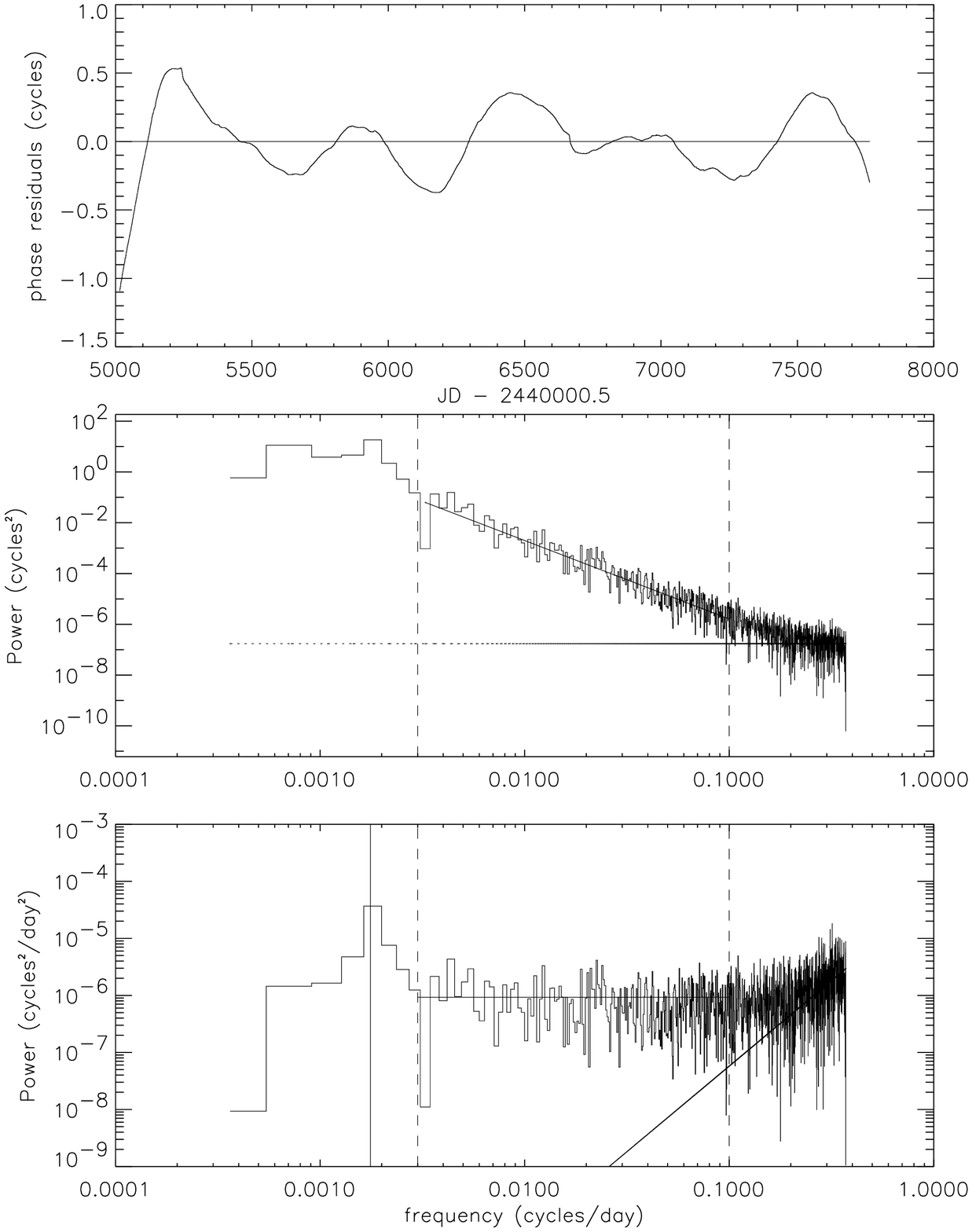,height=6.0in,width=4.6in} }}
\hfill
\parbox{5.5in}{
Fig 6. Top: Crab arrival time residuals obtained from radio observations by
Jodrell Bank after removal of a quartic polynomial, correcting
for dispersion effects {\it and} interpolating onto a uniform grid.
Middle: Hann windowed ($\alpha = 1$) power density
spectrum showing power-law slope of $-3.09 \pm 0.05$ from fit to power
between vertical dashed lines (frequency range $0.003 - 0.1$ $\rm cycles/day$).
Horizontal line shows calculated power level of statistical measurement noise.
Bottom: flattened PDS after multiplication by $(2\pi f)^3$, where f is
frequency. Note the highly significant peak at $f = 0.0018$ $\rm cycles/day$.
Mean power level (between vertical dashed lines) is $9.26(0.50)\times 10^{-7}$
$\rm cycles^2/day^2$. Solid vertical line shows cosine frequency from table 2.
} 

\newpage

\centerline{\hbox{
\psfig{figure=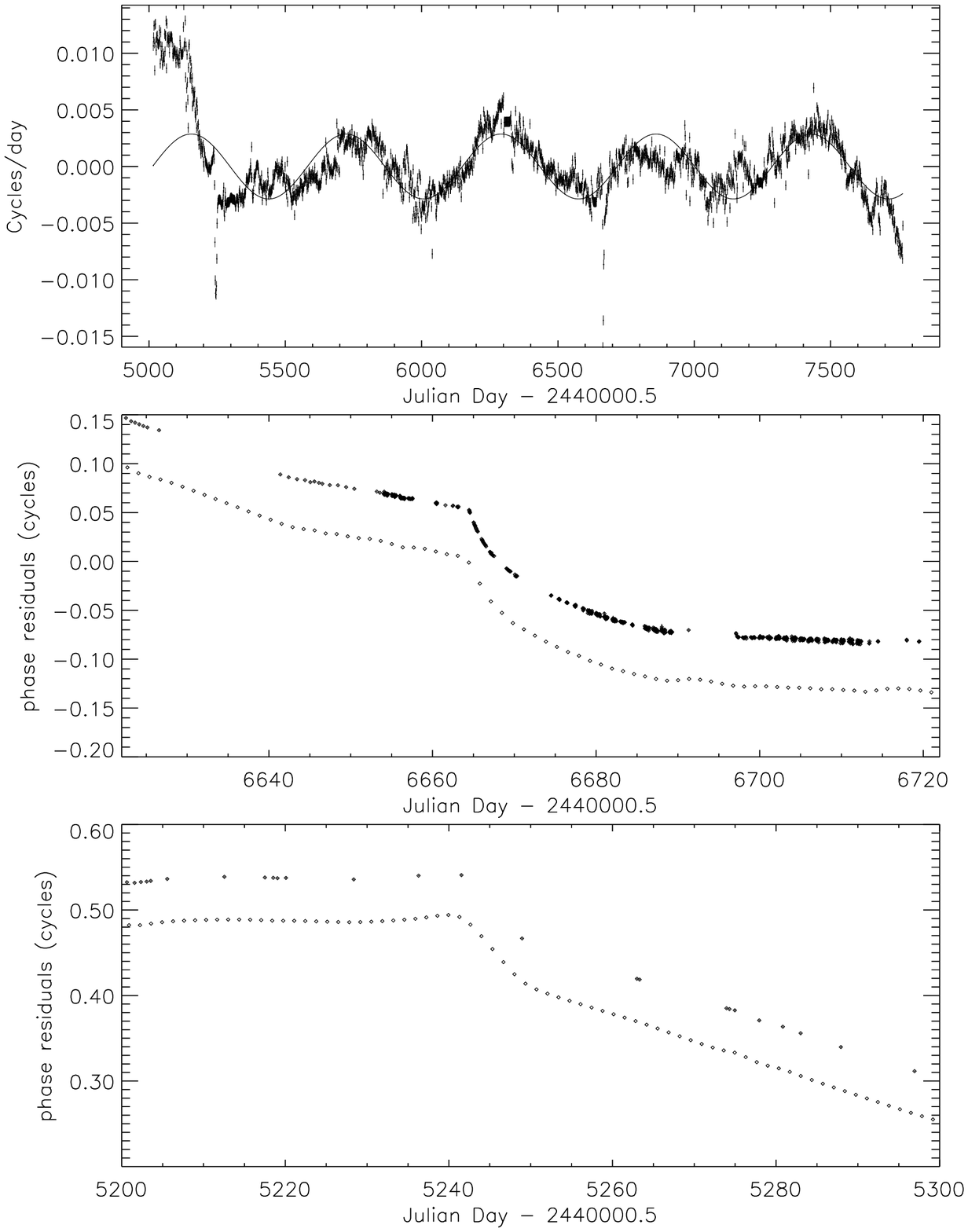,height=6.0in,width=4.6in} }}
\hfill
\parbox{5.5in}{
Fig 7. Top: The numerical 1st derivative of the uniformly binned spline fit to
the Crab timing residuals. The superposed sinusoid is the derivative of
the sinusoid fit to the Crab timing phase residuals. 
Middle: A closeup of the Crab timing residuals at the time of the small
glitch occuring at JD -- 2440000.5 = 6664. The raw mesurements are plotted
in top, underneath (diamonds) is the spline interpolation.  
Bottom: A similar closeup of the region associated with the first large 
excursion in the numerical 1st derivative. A possible glitch may
have occured here.                                                      
}       

\newpage

\centerline{\hbox{
\psfig{figure=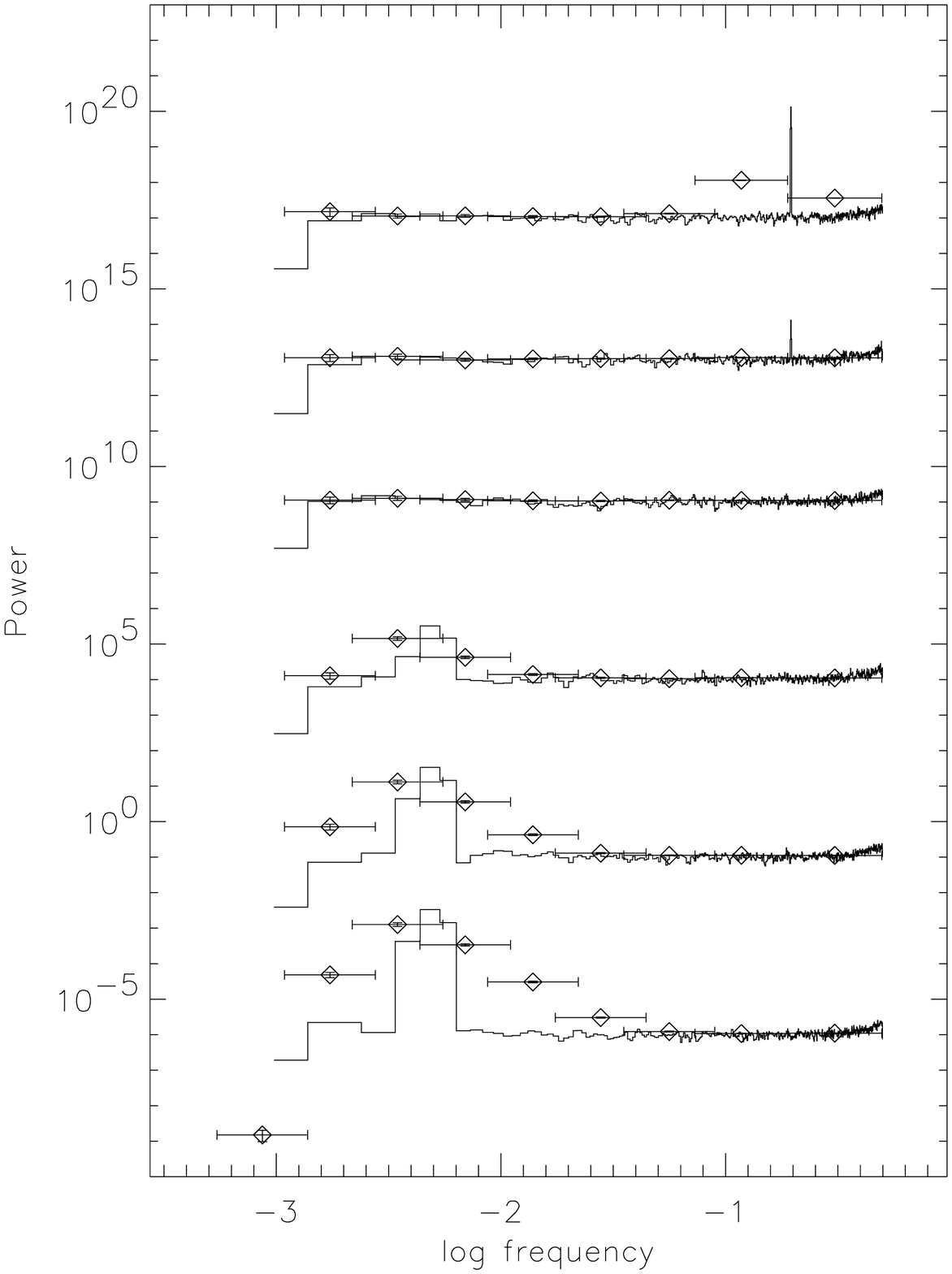,height=6.0in,width=4.6in} }}
\hfill
\parbox{5.5in}{
Fig 8. Comparison of average power spectra computed using the Deeter polynomial method
(diamonds with error bars) and windowed power spectra using a Hann window
with $\alpha=2$ of the same 32 uniformly sampled noise realizations. The noise 
is $1/f^3$ PLN generated using the time domain method and detrended by a cubic 
polynomial fit. The spectra have been flattened for the case of a $1/f^3$ 
power-law. High (top two spectra) and low (bottom three spectra) frequency 
sinusoids have been added to the noise realizations to show the effect of 
discrete features on the power spectrum estimation. The power spectra have 
been shifted in power for display purposes but the noise strength is the same 
in all cases.                                                                  
}       

\newpage

\centerline{\hbox{
\psfig{figure=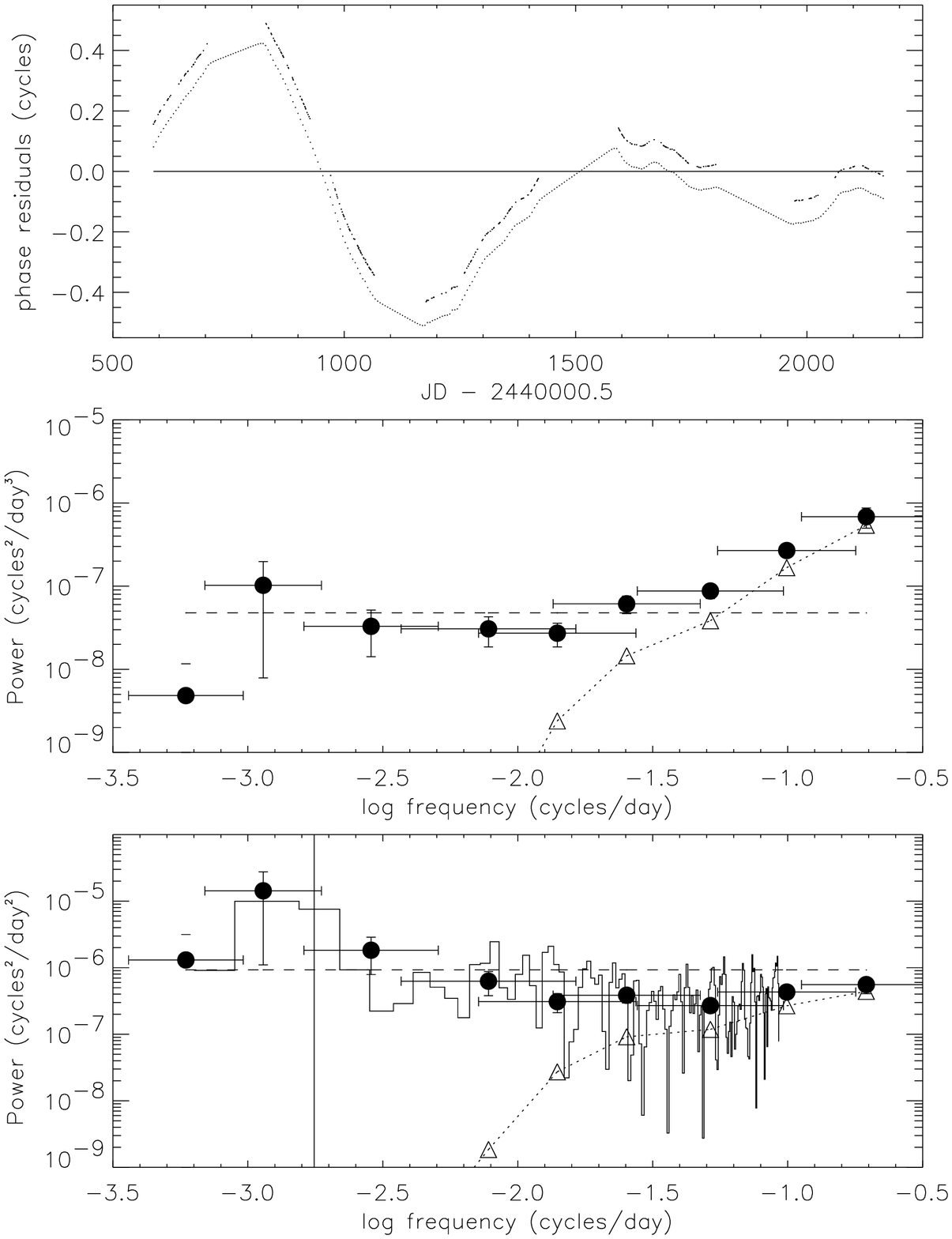,height=6.0in,width=4.6in} }}
\hfill
\parbox{5.5in}{
Fig 9.  Top: Crab pulsar optical timing pulse phase residuals from Groth (1975c)
for period after 1969 glitch. Offset below is a cubic spline interpolation
onto a uniform time series.
Middle: Solid circles mark the Deeter polynomial power spectrum of the pulse 
frequency derivative (flat for $1/f^4$ noise in pulse phase). Triangles 
connected by  dotted line mark computed measurement noise spectrum. 
Dashed line shows power level measured by Deeter (1981) in his analysis
of these data. 
Bottom: Above PDS after multiplication by $(2\pi f)$, which assumes that
the real spectrum is $1/f^3$ noise in pulse phase. Note the ``bump'' in
the power near log frequency $-3.0$. Dashed horizontal line shows noise power 
level measured from Jodrell Bank measurements. Solid vertical line marks
a period of 568 days. Histogram shows Hann windowed power 
($\alpha=2$) power of spline interpolation flattened assuming an power-law 
index of $-3$.        
}

\end{document}